%% file: paper.tex
\documentclass[hyper,11pt,paper]{article}
\usepackage{jheppub}
\usepackage[usenames,dvipsnames,table]{xcolor}
\usepackage{graphicx,amsmath,amssymb,multirow,array,bm,mathrsfs,bbold,bbm}
\usepackage{appendix}
\usepackage{footmisc}
\usepackage{epsf,amsfonts,slashed,soul}
\usepackage{lipsum}
\usepackage{subfig}
\usepackage{pifont}
\usepackage{bbold}
\usepackage{dsfont}
\usepackage{tikz}
\usepackage{xcolor}
\usepackage{verbatim}
\usepackage{placeins}

\usepackage{comment}
\usepackage{simpler-wick}

\allowdisplaybreaks
\newcommand{\bea}{\begin{eqnarray}}
\newcommand{\eea}{\end{eqnarray}}

\input{macros}

\title{Modular chaos, operator algebras, and the Berry phase}

\author[a]{Jan de Boer,}
\author[b,a]{Bahman Najian,}
\author[c,a]{Jeremy van der Heijden,}
\author[d]{and Claire Zukowski}
\affiliation[a]{Institute for Theoretical Physics, University of Amsterdam, Science Park 904, Postbus 94485, 1090 GL Amsterdam, The Netherlands}
\affiliation[b]{Center for Theoretical Physics, Massachusetts Institute of Technology, Cambridge, MA 02139, USA}
\affiliation[c]{Department of Physics and Astronomy, University of British Columbia, 6224 Agricultural Road, Vancouver, B.C. V6T 1Z1, Canada}
\affiliation[d]{Department of Physics and Astronomy, University of Minnesota Duluth,
Duluth, MN 55812, USA}

\emailAdd{J.deBoer@uva.nl}
\emailAdd{bahman@mit.edu}
\emailAdd{jeremy.vanderheijden@ubc.ca}
\emailAdd{czukowsk@d.umn.edu}

\abstract{
Modular Berry transport associates a geometric phase to a zero mode ambiguity in a family of modular operators. In holographic settings, this phase was shown to encode nontrivial information about the emergent spacetime geometry. We reformulate modular Berry transport for arbitrary von Neumann algebras, including giving a precise definition of the zero mode projection in terms of a conditional expectation. For a certain class of state perturbations, we demonstrate that the modular Berry phase gives rise to an emergent symplectic form in the large $N$ limit, extending related results in the context of subregion/subalgebra duality. We also show that the vanishing of the Berry curvature for modular scrambling modes signals the emergence of a local Poincaré algebra, which plays a key role in the quantum ergodic hierarchy. These results provide an intriguing relation between geometric phases, modular chaos and the local structure of spacetime.
}

\begin{document}
\maketitle

\input{introduction}

\input{algebraicberry}

\input{GFF}

\input{OTOC}

\input{discussion}

\section*{Acknowledgements}
It is a pleasure to thank Ramesh Chandra, Mikhail Isachenkov, Nima Lashkari, Antony Speranza, Pim van den Heuvel and Erik Verlinde for discussions. We especially thank Antony Speranza for providing detailed comments on the manuscript. JdB and JvdH are supported by the European Research Council under the European Unions Seventh Framework Programme (FP7/2007-2013), ERC Grant agreement ADG 834878. JvdH also acknowledges support from the National Science and Engineering
Research Council of Canada (NSERC) and the Simons Foundation via a Simons Investigator Award. BN acknowledges support from the American Institute of Physics through the Robert H.G. Helleman Memorial Postdoctoral Fellowship, and was supported by the Spinoza Grant of the Dutch Science Organisation (NWO) awarded to Erik Verlinde. CZ is supported by the National Science Foundation under Award Number 2412608, and acknowledges a Heising-Simons Fellowship as part of the “Observational Signatures of Quantum Gravity” collaboration grant 2021-2818. This work is supported by the Delta ITP consortium, a program of the Netherlands Organisation for Scientific Research (NWO) that is funded by the Dutch Ministry of Education, Culture and Science (OCW).
\appendix

\input{appendix}

\bibliographystyle{JHEP}
\bibliography{vNBerry.bib}

\end{document}

%% file: Macros.tex
\newcommand{\be}{\begin{equation}}
\newcommand{\ee}{\end{equation}}
\newcommand{\beq}{\begin{equation}}
\newcommand{\eeq}{\end{equation}}
\newcommand{\ba}{\begin{eqnarray}}
\newcommand{\ea}{\end{eqnarray}}

\newcommand{\propegator}{{\mathbf{P}}}

%% file: introduction.tex
\section{Introduction}

One of the central ideas driving many of the recent advances in quantum gravity is the notion that spacetime emerges from some underlying quantum mechanical description. The AdS/CFT correspondence provides a concrete example of this idea, where certain features of the bulk anti-de Sitter spacetime can be reconstructed from the boundary conformal field theory. Over the years, research into various aspects of the AdS/CFT correspondence has resulted in a fairly solid understanding of the dictionary between bulk and boundary degrees of freedom. However, despite this explicit realization, the broader mechanism behind the emergence of spacetime remains an open question. 

Recently, it has been proposed that techniques in the theory of operator algebras could shed light on this question. An important aspect of the problem is to understand which elements of the quantum mechanical theory give rise to geometric notions, e.g., subregions, causal structure, and sharp horizons, in the emergent spacetime description. It turns out that the operator algebraic approach is especially effective in revealing features that are amenable to such a spacetime interpretation \cite{DeBoer:2019kdj, Kang:2018xqy,Gesteau:2020rtg,Leutheusser:2021qhd,Leutheusser:2021frk, Chandrasekaran:2022eqq, Leutheusser:2022bgi, deBoer:2022zps,
Furuya:2023fei, 
Gesteau:2023rrx, Ouseph:2023juq, 
Witten:2023xze,
engelhardt:2023xer,
Gesteau:2024dhj, vanderHeijden:2024tdk, Gesteau:2024rpt, Lashkari:2024lkt, Lashkari:2018oke,Gesteau:2021jzp,Crann:2024gkv}. This algebraic version of the holographic duality, in which geometric subregions in the bulk are associated to emergent type III$_1$ von Neumann algebras on the boundary, is known as \emph{subregion/subalgebra duality} \cite{Leutheusser:2022bgi}. In addition, the approach is sufficiently abstract allowing for various applications, both within AdS/CFT and beyond (see, e.g., \cite{Chandrasekaran:2022cip, Penington:2023dql, Kolchmeyer:2023gwa,
Jensen:2023yxy, AliAhmad:2023etg, Kudler-Flam:2023qfl, Kudler-Flam:2023hkl, Faulkner:2024gst, Chen:2024rpx,Kudler-Flam:2024psh,Kolchmeyer:2024fly,Jensen:2024dnl,AliAhmad:2025ukh,Chemissany:2025vye,Speranza:2025joj}).

In the present paper, we focus on one particular aspect of the subregion/subalgebra duality that involves an operator algebraic quantity in the boundary that is analogous to the usual notion of Berry phase in quantum mechanics \cite{Berry:1984jv}. In the original setting, a one-parameter deformation of the theory's Hamiltonian is considered, in such a way that the deformation defines a closed loop in parameter space. Surprisingly, the state of the system does not always return to its initial value, and might acquire a non-trivial phase in the process. Similar ideas have been applied in the context of holography. The idea is to replace the Hamiltonian in the above procedure by the \emph{modular} Hamiltonian in the boundary CFT. Since this involves a parallel transport of modular Hamiltonians, the resulting phase is referred to as the \emph{modular Berry phase} \cite{Czech:2017zfq, Czech:2019vih}. Constructed from a family of modular Hamiltonians, the resulting Berry phase probes the intricate entanglement structure on the boundary, which from the bulk perspective, is represented in the local structure of spacetime.

One can generate a family of modular Hamiltonians in various ways, for example by considering a fixed global state in the boundary CFT and looking at a family of subregions. The corresponding transformations are called \emph{shape-changing}, as they change the shape of the boundary subregion. Their parallel transport is closely related to the notion of \emph{kinematic space} \cite{Czech:2015qta, Czech:2016xec, deBoer:2015kda, deBoer:2016pqk}, and the resulting Berry curvature has been shown to probe bulk lengths and curvatures~\cite{Czech:2017zfq, Czech:2019vih}, also for disconnected regions including islands~\cite{Chen:2022nwf, Aalsma:2024qnf}. Thus, this provides a concrete realization of the idea that entanglement gives rise to geometry. Another class of modular Hamiltonians can be obtained by keeping the subregion fixed, while changing the global state. These deformations, to distinguish them from the shape-varying ones, are called \emph{state-changing} transformations. Recently, an interesting family of such transformations, obtained by acting with Virasoro generators on the global state, has been investigated in the context of AdS$_3$/CFT$_2$ \cite{deBoer:2021zlm}. This discussion has subsequently been generalized to higher dimensions by considering a generic class of coherent state perturbations in the CFT using Euclidean path integral techniques \cite{Czech:2023zmq}. The resulting parallel transport can be carried out explicitly and allows one to reconstruct the phase space of the bulk theory.

While the modular Berry phase has been studied in the context of the AdS/CFT duality, a proper algebraic description in terms of von Neumann algebras is still lacking (see~\cite{Banerjee:2023eew} which discusses algebras for other types of holographic Berry phases~\cite{Nogueira:2021ngh,Banerjee:2022jnv}). The present work aims to bridge this gap. In fact, we will present a framework that is purely algebraic, and is phrased entirely in terms of objects that exist for general von Neumann algebras. This will allow us to define the modular Berry phase, not only in the familiar setting of type I algebras, that is usually discussed in the literature, but also directly in the case of type II and type III von Neumann algebras\footnote{For more background on the type classification of von Neumann algebras we refer to \cite{Witten:2018zxz, Sorce:2023fdx}.}. We will argue that this more general approach can be used to address some interesting physical questions related to the emergence of geometric concepts in gravity. In addition, the framework does not rely on any specifics of the AdS/CFT correspondence, and should therefore, in principle, be more generally applicable.

After setting up the formalism, we will set out to compute the modular Berry phase associated with certain boundary algebras that emerge in the large $N$ limit. Because the corresponding operator algebras constitute type III$_1$ von Neumann algebras, this provides a first nontrivial application of our operator algebraic approach. Technically, the computation involves the von Neumann algebra of generalized free field (GFF) theory \cite{Greenberg:1961mr}, as the $N\to\infty$ limit of the boundary theory is well-described by a GFF theory \cite{Aharony:1999ti, Duetsch:2002hc}. For concreteness, we will phrase most of the discussion in terms of the thermofield double state dual to the eternal black hole geometry, while a generalization to other settings should be possible. We will show that the parallel transport problem for state-changing transformations gives rise to a non-trivial Berry phase, which captures geometric information about the dual bulk spacetime. In fact, we will argue that the modular Berry phase coincides with the bulk covariant phase space symplectic form, therefore directly probing the phase space structure of bulk theory. While a connection between the Berry phase and the bulk symplectic form has been studied before, see, e.g, \cite{Belin:2018fxe, Belin:2018bpg} for the case of pure state Berry phases, or \cite{Kirklin:2019ror, deBoer:2021zlm, Czech:2023zmq} for its modular version, the main result of the present work is a derivation of this statement using properties of the underlying von Neumann algebras.

Another interesting application of our formalism involves the study of modular chaos. The relevant algebraic setup involves so-called \emph{modular scrambling modes}, which saturate the modular chaos bound \cite{DeBoer:2019kdj}, as well as the closely related structure of a \emph{half-sided modular inclusion} of type III von Neumann algebras \cite{Borchers:1991xk, Wiesbrock:1992mg,Borchers:1995zg,Araki-Zsido}. Using the modular scrambling modes, we will construct a one-parameter family of subalgebras, the algebraic equivalent of the shape-changing parallel transport, and compute the associated geometric phase. We will argue that the resulting quantity governs the local structure of spacetime. In particular, we will show that a vanishing modular Berry curvature implies the existence of a local Poincar\'e algebra in the emergent spacetime description. We will comment on the interpretation of this result in the context of the quantum ergodic hierarchy \cite{Furuya:2023fei, Gesteau:2023rrx, Ouseph:2023juq}. We will further strengthen the connection with modular chaos by writing down a modular version of the out-of-time-ordered correlator (OTOC) in terms of a suitable geometric phase. Its exponential growth with maximal Lyapunov coefficient is a signature of maximal modular chaos.\\

\noindent {\bf Outline:} The rest of this paper is organized as follows. In Section \ref{sec:algebraBerry}, we present the modular Berry phase within the framework of von Neumann algebras. After introducing the notion of modular flow and modular zero modes, we present a definition for the modular Berry curvature. It relies on extracting the modular zero mode component from an operator using a projection map, which we define algebraically in terms of a conditional expectation. We then show that the vanishing of the Berry curvature distinguishes different types of von Neumann algebras. In Section \ref{sec:symform}, we consider the case of the thermofield double state in the large $N$ limit and compute the modular Berry curvature associated to a class of state-changing unitaries. Next, we give a bulk description using the extrapolate dictionary and the covariant phase space formalism, and find that the modular Berry curvature agrees with the bulk symplectic form. Section \ref{sec:modchaos} explores the modular Berry curvature as a tool to probe the connection between the modular structure and quantum chaos in terms of shape-changing deformations. We utilize the concept of half-sided modular inclusion to show that the vanishing of the modular Berry curvature for modular scrambling modes leads to an emergent Poincaré algebra.
Finally, we define a modular version of the OTOC and show that it can be computed as a Berry phase for appropriately chosen global perturbations. We conclude with a discussion of some subtleties and propose directions for future research.

%% file: algebraicberry.tex
\section{An algebraic version of the modular Berry phase}
\label{sec:algebraBerry}

In this section, we will introduce the modular Berry phase in the language of von Neumann algebras. The construction relies mostly on some ingredients of the modular theory of Tomita and Takesaki \cite{Takesaki:1970aki}. While we try to be somewhat self-contained, the uninitiated reader might find it useful to consult some background literature on von Neumann algebras \cite{takesakiII,Witten:2018zxz}, where aspects of the modular theory are reviewed in more detail.

The starting point of the construction is a von Neumann algebra of operators $\mathcal{A}$ acting on a Hilbert space $\mathcal{H}$. One can consider, for example, a CFT Hilbert space $\mathcal{H}$ and the algebra $\mathcal{A}$ being the von Neumann algebra generated by the operators localized in a subregion $A$. Given a global state $|\psi\rangle \in \mathcal{H}$, it is common to consider the reduced matrix $\rho_{A}$ associated to the subregion $A$ by tracing out the complement subregion $\bar{A}$. This is usually done by assuming a tensor factorization Hilbert space, $\mathcal{H}=\mathcal{H}_A\otimes \mathcal{H}_{\bar{A}}$, and defining $\rho_A$ via a partial trace. It turns out that for general von Neumann algebras such a tensor factorization may not exist, for instance when dealing with algebras associated to local subregions in a CFT. Instead, one needs to consider the so-called \emph{modular operator} $\Delta_{\psi}$, which in the case of type I algebras is defined as  $\Delta_{\psi} = \rho_A \otimes \rho_{\bar{A}}^{-1}$, where $\rho_{\bar{A}}$ denotes the reduced density matrix of the complement subregion. Note that $\Delta_{\psi}$ acts on both the operators in region $A$ as well as in the complement $\bar{A}$. One of the main ideas of the Tomita-Takesaki theory is that the modular operator still makes sense as an (unbounded) operator on the Hilbert space, even in situations where the  density matrices, $\rho_{A}$ and $\rho_{\bar{A}}$, do not exist. 

The modular operator contains information on the entanglement structure of the global state $|\psi\rangle$. One way of probing this is by considering a family of modular operators $\Delta_{\psi}(\lambda)$ parametrized by some external parameter $\lambda$. There are various ways to obtain such deformations. Note that the modular operator depends both on the algebra and the state that we choose. An important class of deformations involves changing the global state, while keeping the algebra $\mathcal{A}$ fixed. Following \cite{Summers:2003tf}\footnote{We thank Nima Lashkari for pointing this reference out to us.}, we will refer to this as the ``two states, one algebra'' scenario. Let us consider perturbations of the form
\beq 
|\psi(\lambda)\rangle = u(\lambda)|\psi\rangle~,
\eeq
where we excite the state with some one-parameter family of unitaries $\lambda \mapsto u(\lambda)\in \mathcal{A}$. The corresponding family of modular Hamiltonians is given by 
\beq \label{eq:uDelta}
\Delta_{\psi}(\lambda) = u(\lambda) \Delta_{\psi} u(\lambda)^{\dagger}~.
\eeq
Equation \eqref{eq:uDelta} defines a path in the space of modular operators. However, from the perspective of the algebra $\mathcal{A}$, not all perturbations lead to physically distinct modular operators. To be precise, if we use a unitary $\lambda \mapsto v(\lambda)\in \mathcal{A}$ with the property that $\Delta_{\psi} v(\lambda) = v(\lambda) \Delta_{\psi}$, the resulting modular operator does not change. As a result, we have a ``frame'' ambiguity along the path in the sense that we can send $u(\lambda)\to u(\lambda)v(\lambda)$ without changing the modular operator at that point. The operators $v(\lambda)$ that have this property are called \emph{modular zero modes}. In the following, we will denote the subalgebra of modular zero modes by $\mathcal{A}^{\psi}$. A precise definition will follow momentarily. While the modular zero modes described above are operators in $\mathcal{A}$, there is also a notion of global zero mode in $B(\mathcal{H})$, which we will discuss in Section \ref{sec:modchaos}.

The zero mode frame ambiguity can be expressed most clearly in terms of an auxiliary gauge field. To define it, one needs a way to project an arbitrary perturbation onto its zero mode piece. We will argue that one needs to define a conditional expectation $\mathcal{E}_{0}:\mathcal{A}\to \mathcal{A}^{\psi}$, which maps the algebra onto the subalgebra of modular zero modes, and a corresponding zero mode projector $P_0$ on the level of the generators of the algebra. In the absence of an inner product, the zero mode projector $P_0$ is not a priori unambiguously defined (see, e.g., Appendix C of \cite{deBoer:2021zlm}). One of the main points of this paper is to give a precise definition of the zero mode projector within the algebraic framework. Given $P_0$, one then defines the following connection on the space of parameters
\beq \label{eq:gaugefieldA}
\Gamma = P_0(u(\lambda)^{\dagger}\delta u(\lambda))~.
\eeq
Note that \eqref{eq:gaugefieldA} indeed has the correct transformation properties of a gauge field. If we introduce a change of zero mode frame $u(\lambda)\to u(\lambda)v(\lambda)$ with $v(\lambda)\in \mathcal{A}^{\psi}$, the gauge field transforms as  
\beq 
\Gamma \to v(\lambda)^{\dagger}\Gamma v(\lambda)+v(\lambda)^{\dagger}\delta v(\lambda)~,
\eeq
where the analogue of the ``gauge group'' is the algebra of modular zero modes. Now that we have defined an auxiliary gauge field on the space of deformations, it is natural to ask when it exhibits a non-trivial curvature. If we denote the generators of two deformations by $X_1$ and $X_2$, respectively, with the additional requirement that $P_0(X_1)=P_0(X_2)=0$, it was shown that the corresponding curvature can be computed using the formula (see Appendix B of \cite{deBoer:2021zlm} for a derivation\footnote{While the proof in \cite{deBoer:2021zlm} is phrased in terms of Lie algebras, the same reasoning should hold in the present context. The main non-trivial assumption that goes into the derivation is the existence of a zero mode projector and the fact that any generator $Y$ has a unique decomposition $Y=Y_0+I$ as a sum of a zero mode piece $Y_0$ and a non-zero mode piece $I$.}):
\beq \label{eq:F=P0}
F = P_0([X_1,X_2])~.
\eeq
Because of the similarity with the usual Berry phase for pure states and the fact that we are using the modular structure of the theory to define it, the operator in \eqref{eq:F=P0} is called the \emph{modular Berry curvature}. The main achievement of the present work is to set up the formalism that is applicable to cases where the underlying operator algebra is some abstract von Neumann algebra. In particular, we give an algebraic definition of the zero mode projector $P_0$. Now that we have outlined the general prescription, we will use the rest of this section to work out some of the details.

\subsection{Modular flow and modular zero modes}

As before, we let $\mathcal{A}$ be a von Neumann algebra acting on a Hilbert space $\mathcal{H}$, and consider a global state $|\psi \rangle \in \mathcal{H}$. We require $|\psi\rangle$ to be cyclic and separating. The cyclic property means that one can generate all the states in the Hilbert space $\mathcal{H}$ by acting on $|\psi\rangle$ with operators in the algebra $\mathcal{A}$, and the state is called separating if it cannot be annihilated by any operator in $\mathcal{A}$. Physically, the cyclic and separating property of the state can be understood as a manifestation of its high degree of entanglement. At times, it will be useful to think about states as linear functionals that arise from taking expectations values
\beq
\psi: \mathcal{A}\to \mathbb{C}: a\to \langle \psi |a|\psi \rangle~.
\eeq 
Given a cyclic and separating state, one can use the modular theory to define a natural action on the algebra, which is called \emph{modular flow}. It provides the algebra with an internal ``clock'' arising from entanglement with external degrees of freedom. The modular flow is defined in terms of the modular operator. To define it, we first introduce the so-called Tomita map $S_{\psi}$. It implements the operation of taking the Hermitian conjugate on the Hilbert space via the assignment
\beq \label{eq:Spsi}
S_{\psi} a|\psi\rangle = a^{\dagger}|\psi\rangle~.
\eeq
This is then extended to the full Hilbert space by anti-linearity. The cyclic and separating property of the state $|\psi\rangle$ ensures that this leads to a well-defined map. Although the Tomita map $S_{\psi}$ is generally unbounded, it still has some favorable properties. For example, it is possible to define the adjoint map $S_{\psi}^{\dagger}$ up to some subtleties with the respective domains of these maps. The modular operator is now defined by setting
\beq 
\Delta_{\psi}
\equiv S_{\psi}^{\dagger}S_{\psi}~.
\eeq
By construction, it is a positive Hermitian operator. For later use, we note that it follows from \eqref{eq:Spsi} that the modular operator leaves the state $|\psi\rangle$ invariant:
\beq
\Delta_{\psi}|\psi\rangle =|\psi\rangle~.
\eeq
While the modular operator is in general unbounded, we can consider a family of bounded unitaries $\Delta^{is}$ for $s\in \mathbb{R}$ constructed from it. It is a highly nontrivial fact of the Tomita-Takesaki theory that the map
\beq \label{eq:modflow}
\sigma_s(a) \equiv \Delta_{\psi}^{-is}a\Delta_{\psi}^{is}~, \hspace{20pt} s\in \mathbb{R}~,
\eeq
defines a one-parameter group of automorphisms of the algebra $\mathcal{A}$. The fact that $\sigma_s$ maps the algebra to itself is one of the main theorems of the modular theory. The group of automorphisms in \eqref{eq:modflow} is called the modular flow. Physically, it provides the algebra with some internal notion of time, and for this reason, the coordinate $s$ is referred to as \emph{modular time}. In the special case that we have algebras that admit irreducible representations\footnote{This corresponds to the usual case in quantum mechanics where we consider a factorized Hilbert space $\mathcal{H}=\mathcal{H}_A\otimes \mathcal{H}_{\bar{A}}$ with algebra $\mathcal{A}=B(\mathcal{H}_A)$ and commutant $\mathcal{A}'=B(\mathcal{H}_{\bar{A}})$.} the modular operator can be factorized in terms of density matrices and the modular flow is given by conjugating with $\rho_{\psi}^{is}\in \mathcal{A}$. An automorphism that can be implemented by conjugation with an element of the algebra is called \emph{inner}. If the opposite is true, it is called an \emph{outer} automorphism. An important result in the classification of von Neumann algebras is that for type I and type II algebras the modular flow is inner, while for type III algebras it is outer.

Now, we would like to focus on a particular subalgebra of operators that are unchanged by the modular flow. The invariant subalgebra is given by operators that satisfy
\beq 
\sigma_s(a)=a~, \qquad \forall s\in \mathbb{R}~.
\eeq 
Operators that satisfy this condition are called \emph{modular zero modes}. Because the modular flow defines an automorphism of the algebra, the modular zero modes naturally form a subalgbra of $\mathcal{A}$. This subalgebra is also known as the \emph{centralizer}, and denoted by $\mathcal{A}^{\psi}$. One can think of the zero modes as being the analogue of conserved charges for the modular operator: If we interpret the modular flow as being a time flow of some Hamiltonian, the so-called \emph{modular Hamiltonian} defined by 
\beq 
K_{\psi} \equiv -\log \Delta_{\psi}~,
\eeq 
the modular zero modes are precisely the operators in the algebra that do not evolve with time, and are therefore conserved. Indeed, by taking a derivative with respect to $s$, one sees that the modular zero modes are precisely those operators that commute with $K_{\psi}$:
\beq \label{eq:commuteK}
-i\frac{d}{ds}\Big|_{s=0} \sigma_s(a) = [K_{\psi}, a] = 0~.
\eeq

\subsubsection{A characterization of the centralizer}

It is useful to introduce a slightly different perspective on the subalgebra of modular zero modes. To this end, we first define the \emph{commutant} $\mathcal{A}'$ of $\mathcal{A}$ which is the von Neumann algebra of all operators in $B(\mathcal{H})$ that commute with all operators in $\mathcal{A}$. The \emph{center} of the algebra $\mathcal{A}$ is now given by the intersection
\beq 
Z(\mathcal{A})=\mathcal{A}\cap \mathcal{A}'~.
\eeq
The algebra $\mathcal{A}$ is called a \emph{factor} when the center is trivial, $Z(\mathcal{A})=\mathbb{C}\cdot \mathds{1}$. Therefore, in a factor there are no operators in $\mathcal{A}$ that commute with all operators in $\mathcal{A}$. However, there can be operators that satisfy this condition within expectation values. To be precise, we consider the operators $a\in \mathcal{A}$ with the property that
\beq \label{eq:conditioncentralizer}
\psi([a,b]) = \langle \psi | [a,b] |\psi \rangle = 0~, \qquad \forall b \in \mathcal{A}~.
\eeq 
Clearly, operators in the center $Z(\mathcal{A})$ satisfy the condition in  \eqref{eq:conditioncentralizer}. We will now show that there is a one-to-one correspondence between operators satisfying \eqref{eq:conditioncentralizer} and the modular zero modes:
\begin{equation} \label{eq:1-1zeromodes}
\psi([a,b])= 0~, \quad \forall b \in \mathcal{A} \qquad \longleftrightarrow \qquad \sigma_s(a)=a~, \quad \forall s\in \mathbb{R}~.
\end{equation}
The above correspondence will be useful in understanding the structure of the algebra of modular zero modes. To be somewhat self-contained, we provide a short argument for this statement (see also \cite[Theorem VIII.2.6]{takesakiII} and \cite[Lemma 12]{Furuya:2023fei}).

To prove \eqref{eq:1-1zeromodes}, we need to argue for both implications. Let us first consider an operator $a\in \mathcal{A}$ that is fixed by the modular flow, and show that it satisfies \eqref{eq:conditioncentralizer}. By analyticity\footnote{The subalgebra  of elements on which the modular flow acts entirely analytic is called the \emph{Tomita subalgebra}, denoted by $\mathcal{A}_{\rm T}$. One can show that $\mathcal{A}_{\rm T}$ is dense in $\mathcal{A}$ with respect to the strong operator topology (an operator is considered to be small in the ``strong'' operator topology when its action on states is small: We say that $a_n\to a$ converges if $(a-a_n)|\psi\rangle \to 0$ for every $|\psi\rangle\in \mathcal{H}$). For this reason, one can assume without loss of generality that the function $\sigma_s(a)$ is analytic.\label{foornote:4}}, we can extend the action of modular flow to imaginary values of modular time, which implies that $\sigma_{s=-i}(a)=\Delta_{\psi}^{-1}a\Delta_{\psi}=a$. Hence, the operator $a$ commutes with the modular operator $\Delta_{\psi}$ associated to the state $\psi$. We will now use the KMS condition, which states that for two operators $a,b\in \mathcal{A}$ the following condition holds
\beq \label{eq:KMS}
\langle \psi |ba | \psi \rangle =\langle \psi |a \Delta_{\psi} b |\psi \rangle~.
\eeq 
The KMS condition \eqref{eq:KMS} combined with $\Delta_{\psi}|\psi\rangle=|\psi\rangle$ now implies that  
\beq 
\psi(ba)=\langle \psi |a \Delta_{\psi} b |\psi \rangle = \langle \psi |\Delta_{\psi}ab |\psi \rangle = \psi(ab)~,
\qquad \forall b\in \mathcal{A}~,
\eeq
which shows that $a$ satisfies \eqref{eq:conditioncentralizer}. This proves the implication (right) $\to$ (left) in \eqref{eq:1-1zeromodes}. Conversely, suppose that $a$ satisfies \eqref{eq:conditioncentralizer}. We take $b \in \mathcal{A}$ to be an analytic element for the modular flow (see footnote \ref{foornote:4}) and consider the function
\beq 
\mathcal{F}(s)\equiv\psi(b\sigma_s(a))~.
\eeq
One can show that $\mathcal{F}$ naturally extends to an analytic function on the strip that $-1< \mathfrak{Im}\, s < 0$ which is bounded on the closure of the strip \cite{Witten:2018zxz}. Using that the modular flow associated to $\psi$ leaves the state invariant, $\psi \circ \sigma_s = \psi$, it follows that
\beq \label{eq:someeq}
\psi(\sigma_s(a)b)=\psi(a\sigma_{-s}(b))=\psi(\sigma_{-s}(b)a)=\psi(b\sigma_s(a))~,
\eeq
for $s\in \mathbb{R}$. At the second equality, we have used \eqref{eq:conditioncentralizer}. Using the KMS condition \eqref{eq:KMS}, we can write the left-hand side of \eqref{eq:someeq} also as
\beq \label{eq:someeq2}
\psi(\sigma_s(a)b)= \langle\psi |b \Delta_{\psi} \sigma_s(a) |\psi\rangle =\psi(b\sigma_{s+i}(a))~.
\eeq
Combing \eqref{eq:someeq} and \eqref{eq:someeq2} shows that the function $\mathcal{F}$ satisfies the periodicity condition $\mathcal{F}(s)=\mathcal{F}(s+i)$ for all $s\in \mathbb{R}$. An analytic function is determined by its restriction to the real line $\mathbb{R}$, so in particular we can extend the identity $\mathcal{F}(s)=\mathcal{F}(s+i)$ to complex values of $s$. Combining this periodicity property with boundedness on the strip, it follows that $\mathcal{F}$ is an entire bounded function on the complex plane. By Liouville's theorem, $\mathcal{F}$ must be constant. As a consequence, we have 
\beq 
\langle \psi |b\sigma_{s}(a)|\psi \rangle=\mathcal{F}(s)=\mathcal{F}(0)=\langle \psi| b a |\psi \rangle~.
\eeq
Since the states $b|\psi\rangle$ with $b$ analytic for the modular flow form a dense subset of $\mathcal{H}$, it follows that 
\beq 
\sigma_s(a)|\psi\rangle = a|\psi\rangle~.
\eeq
Using the separating property of $|\psi\rangle$, we now have $\sigma_s(a)=a$ for all $s\in \mathbb{R}$ as required. This proves the implication (left) $\to$ (right) in \eqref{eq:1-1zeromodes}. We conclude that the algebra of modular zero modes is characterized by the condition \eqref{eq:conditioncentralizer}. 

\subsubsection{Modular zero modes and algebra type}

\label{sec:modularzeromodes}

Now that we have given two different descriptions of the modular zero mode algebra, we can study its structure in a bit more detail. As mentioned before the modular operator is a two-sided operator, that is generally unbounded. For this reason, it is not part of the space of zero modes $\mathcal{A}^{\psi}$ as we have defined it. However, when the modular flow is an \emph{inner} automorphism, we can identify a unitary operator in the algebra $\mathcal{A}$ that implements the modular flow. As we will explain below, this unitary will be a non-trivial modular zero mode, that plays the role of the modular operator. We will use this fact to say something about the algebra type of $\mathcal{A}$. 

Note that the identity operator $\mathds{1}$ is always a modular zero mode by definition. We will now argue that when this is the only modular zero mode, the algebra $\mathcal{A}$ must be a type III factor. To investigate this, we assume that the modular zero mode space is trivial in the sense that 
\beq \label{eq:assumptionApsi}
\mathcal{A}^{\psi} = \mathbb{C}\cdot \mathds{1}~.
\eeq
Since $Z(\mathcal{A})\subset \mathcal{A}^{\psi}$, the algebra $\mathcal{A}$ is automatically a factor. Suppose now that the modular flow is an inner automorphism, which means that there exists some family of unitaries $u_{s}\in \mathcal{A}$ that implements the action of modular flow $\sigma_s$ by conjugation
\beq \label{eq:inner}
\sigma_{s}(a)=u_{s}au_{s}^{\dagger}~.
\eeq
One can then use \eqref{eq:inner} together with the fact that $\psi \circ \sigma_s = \psi$ to show that 
\beq 
\psi(au_{s})=\psi(\sigma_{s}(au_{s}))=\psi(u_{s}au_{s}u_{s}^{\dagger}) = \psi(u_{s}a)~.
\eeq
Hence, the unitary $u_{s}$ is part of the centralizer of the state $\psi$, and thus a modular zero mode. By assumption, we only have trivial modular zero modes \eqref{eq:assumptionApsi}, so $u_{s}$ must be a complex phase. It follows that the modular flow acts trivially on all operators $a\in \mathcal{A}$,
\beq 
\sigma_{s}(a)=a~,
\eeq
which shows that $\mathcal{A}\subset \mathcal{A}^{\psi}$, and consequently $\mathcal{A}=\mathbb{C}\cdot \mathds{1}$. We conclude that a non-trivial von Neumann algebra with no non-trivial zero modes cannot have a modular flow that is inner for all times, and therefore it is a type III factor. This shows that type I and type II von Neumann algebra always have non-trivial modular zero modes, as there is always a family of unitary zero modes that implement the modular flow. One can actually prove a slightly stronger statement (see proof of \cite[Theorem 3]{Longo:1979dw}), namely that the algebra must be type III$_1$ assuming it is not trivial. To summarize:
\begin{quotation}
\noindent \textit{In the absence of non-trivial modular zero modes, a non-trivial von Neumann algebra must be a type III$_1$ factor.}
\end{quotation}
The converse of the above statement does not hold. Type III$_1$ von Neumann algebras can, and in general do, have states which exhibit non-trivial modular zero modes, although we are not aware of a simple physical setup in which this situation arises. We refer to \cite{2023arXiv230514217M} for a discussion of states on type III$_1$ algebras with non-trivial centralizer.

The zero mode ambiguity that underlies the definition of the modular Berry phase heavily depends on which modular zero modes are present. For type III algebras there is always the identity zero mode, but in order to have nontrivial zero mode contributions, for example coming from the modular operator itself, one needs to enlarge the centralizer. One way of achieving this technically is by computing the so-called \emph{crossed product} of the algebra \cite{Witten:2021unn} with a group constructed from the modular zero mode flows (see \cite{AliAhmad:2024eun} for a discussion of crossed products with respect to locally compact groups containing the modular flow). Roughly speaking, the crossed product amounts to adding a generator to the algebra that implements the action of zero mode flow, which by construction defines an element of the centralizer. Physically, the procedure has been interpreted in certain cases in terms of adding an observer to the physical system (see, e.g., \cite{Chandrasekaran:2022cip,Jensen:2023yxy,Witten:2023xze,DeVuyst:2024pop,Speranza:2025joj}). One can take a similar perspective here, in the sense that to properly define the notion of a modular zero mode in the algebraic setting, one requires the notion of an observer that can keep track of this quantity. We will come back to this point in the discussion section.

\subsubsection{A projection onto the zero mode algebra}

To define the modular Berry phase one needs a procedure for extracting the modular zero mode piece from an arbitrary operator. As we will now explain, this requires a careful treatment of the algebra of modular zero modes. First, we will give an abstract definition of the projection map in terms of a conditional expectation, and then provide an explicit expression under certain additional assumptions. In particular, we will retrieve the zero mode projection that was used before in \cite{Czech:2023zmq}.

An important property of the modular zero modes is that they are left invariant by the modular flow. Indeed, by definition we have 
\beq 
\sigma_s(\mathcal{A}^{\psi})=\mathcal{A}^{\psi}~, \quad \forall s\in \mathbb{R}~.
\eeq
Given a subalgebra that is invariant under the modular flow, we can invoke a theorem by Takesaki \cite{takesaki1972conditional} to show that there exists a map $\mathcal{E}_{0}:\mathcal{A}\to \mathcal{A}^{\psi}$ that acts like a projection on the level of the algebras. The map $\mathcal{E}_0$ is called a \emph{conditional expectation}, and satisfies the following bimodule property:
\beq \label{eq:bimoduleproperty}
\mathcal{E}_0(a_1ba_2)=a_1\mathcal{E}_0(b)a_2~, \qquad a_1, a_2 \in \mathcal{A}^{\psi}~, b\in \mathcal{A}~.
\eeq 
There are some additional technical requirements that the state $\psi$ needs to satisfy in order for the theorem to apply. A state $\phi$ is called \emph{normal} if it is continuous in the $\sigma$-weak topology\footnote{There are different topologies that can be used to define a notion of continuity for states. The \emph{weak topology} amounts to taking limits within the algebra where a sequence $a_n\to a$ converges if all of the expectations values $\langle \xi | a_n |\xi\rangle \to \langle \xi | a |\xi\rangle$ converge for arbitrary states $\xi$. For the \emph{$\sigma$-weak topology} one additionally needs to check convergence for all expectation values with mixed states as well, i.e., $\mathrm{tr}(\rho a_n)\to \mathrm{tr}(\rho a)$ for all trace-class operators $\rho$. See \cite{Sorce:2023gio} for more details. A state $\phi$ is normal if $\phi(a_n)\to \phi(a)$ when $a_n\to a$ in the $\sigma$-weak topology. \label{footnote:topologies}}, 
and we say that the state $\phi$ is \emph{faithful} when $\phi(a^{\dagger}a)=0$ implies $a=0$. Given a faithful, normal  state $\psi$ on the algebra $\mathcal{A}$,
it now follows from Takesaki's theorem that the invariance of $\mathcal{A}^{\psi}$ under the modular flow is equivalent to the existence of a faithful, normal conditional expectation
\beq \label{eq:condexp0}
\mathcal{E}_0:\mathcal{A}\to \mathcal{A}^{\psi}~,
\eeq
with the property that it leaves the state $\psi$ invariant, 
\beq \label{eq:psiE0}
\psi\circ \mathcal{E}_0 = \psi~.
\eeq
In fact, this is the unique map satisfying these condition. The conditional expectation \eqref{eq:condexp0} provides a natural projection map onto the subalgebra of modular zero modes, and we will use it to define the modular Berry curvature. Conditional expectations have appeared before in the algebraic approach towards holography, for example, to define the holographic map between bulk and boundary degrees of freedom \cite{Faulkner:2020hzi, Furuya:2020tzv, Gesteau:2021jzp, AliAhmad:2024saq, AliAhmad:2025oli} or to discuss black hole information in the algebraic context \cite{vanderHeijden:2024tdk}. In the present paper, it is used to set up a notion of parallel transport for modular operators. 

The definition of the projection map in terms of a conditional expectation is rather abstract. To make a connection with previous work, we show that \eqref{eq:condexp0} agrees with the modular zero mode projection as previously defined (see, e.g., \cite{Czech:2023zmq}) by writing out an explicit expression for $\mathcal{E}_0$ under some assumptions. For example, we can consider a situation where it is possible to ``diagonalize'' the modular flow in some appropriate sense, so that we can define a suitable notion of ``eigenspace'' for the modular flow $\mathcal{A}_{\omega}$ consisting of all operators $a\in \mathcal{A}$ that satisfy
\beq \label{eq:eigenvalue}
\sigma_s(a) = e^{i\omega s} a~,
\eeq  
for all $s \in \mathbb{R}$. A priori, the ``eigenvalues'' $\omega$ can be any complex number, but some of the eigenspaces could be empty depending on the state $\psi$. Note that the eigenspace $\mathcal{A}_{\omega}$ is clearly a linear space, but in general not a subalgebra. To see this, we first observe that $\sigma_s(a^{\dagger})=e^{-i\omega^* s} a^{\dagger}$, which follows from taking the Hermitian conjugate of \eqref{eq:eigenvalue}. As a consequence, we have $\mathcal{A}_{\omega}^{\dagger} = \mathcal{A}_{\omega^*}$, which shows that $\mathcal{A}_{\omega}$ is closed under taking the Hermitian conjugation when $\omega\in \mathbb{R}$. Moreover, we consider $a_{\omega}\in \mathcal{A}_{\omega}, b_{\omega'} \in \mathcal{A}_{\omega'}$, and note that  
\beq 
\sigma_s(a_{\omega}b_{\omega'}) = \sigma_s(a_{\omega})\sigma_s(b_{\omega'}) = e^{i(\omega+\omega')s} a_{\omega}b_{\omega'}~,
\eeq
which shows that the algebra $\mathcal{A}_{\omega}$ is not closed under multiplication, unless $\omega=0$. In the latter case, we see from \eqref{eq:eigenvalue} that the eigenspace $\mathcal{A}_{\omega=0}$ coincides with the subalgebra of modular zero modes. If the linear span of all non-zero eigenspaces $\mathcal{A}_{\omega}$ is dense within the algebra $\mathcal{A}$\footnote{This happens when a faithful normal state $\psi$ is \emph{almost periodic} which means that the corresponding modular operator $\Delta_{\psi}$ has a total set of eigenvectors \cite[Definition 3.7.1]{Connes1973}.}, we have a well-defined notion of modular Fourier decomposition that allows one to decompose an arbitrary operator in terms of modes $a_{\omega}$ on which the action of modular flow is simple, namely via multiplication by a phase $e^{i\omega s}$. The infinitesimal version of \eqref{eq:eigenvalue} is given by 
\beq 
[K_{\psi
},a_{\omega}] = -i \frac{d}{ds}\Big|_{s=0} \sigma_s(a_{\omega}) = -i \frac{d}{ds}\Big|_{s=0} e^{i\omega s} a_{\omega} = \omega a_{\omega}~,
\eeq
and therefore, the modular Fourier modes $a_{\omega}$ are precisely the eigenoperators of the adjoint action of $K_{\psi}$ with eigenvalue $\omega$. This is the way they were introduced in \cite{Czech:2023zmq} in order to compute an explicit expression for the modular Berry curvature. Similarly, eigenoperators of the adjoint action of the modular Hamiltonian were discussed before in the context of two-dimensional CFT \cite{deBoer:2021zlm}, where they are explicitly realized in terms of a continuous version of the Virasoro algebra.

Note that for general states the modular eigenoperators in \eqref{eq:eigenvalue} can be somewhat singular objects. This is related to the fact that in type I algebras, the intersection of the spectrum of all modular operators is always a point spectrum, so that an eigenvalue equation of the form \eqref{eq:eigenvalue} makes sense, whereas in type III$_1$ algebras the intersection of the spectrum of all modular operators does not include any point spectrum. As described above, one situation for which the modular Fourier modes make sense is when the modular operator is diagonalizable, in which case we expect the space of modular zero modes to be non-trivial. The other case where you have well-defined eigenoperators is when working with integrable weights\footnote{A \emph{weight} $\psi$ is a linear functional which generalizes the notion of a state in the sense that it can take on the value $\infty$ on some operators. We say that a weight is \emph{semi-finite} if the set of operators $a$ with $\psi(a^{\dagger}a)<\infty$ is dense in the full algebra $\mathcal{A}$ with respect to the $\sigma$-weak topology. For \emph{integrable} weights the corresponding modular automorphism is integrable in an appropriate sense (we refer to \cite{takesakiII,ConnesTakesakiflow} for a precise definition).} (see, e.g., \cite{Speranza:2025joj}). The weight in that case is not semi-finite on the eigenoperators, so they do not lead to normalizable eigenstates for the modular operator.

A natural guess for extracting the zero mode piece is to simply average the operator over all modular times. If we introduce the correct normalization, it turns out that such an averaging procedure indeed defines a conditional expectation with the right properties. Formally, we can write down the following expression \cite{haagerup1979operatorII}:
\beq \label{eq:zeromodeprojector}
\mathcal{E}_{0}(a)\equiv\lim_{\Lambda \to \infty, \mathrm{weak}}\frac{1}{2\Lambda}\int_{-\Lambda}^{\Lambda}ds\,\sigma_s(a)~,
\eeq
where the limit is taken with respect to the weak operator topology\footnote{To define a projection operator that acts directly on the Hilbert space, one actually requires convergence in the strong operator topology. In present setting, this statement is closely related to von Neumann's ergodic theorem \cite{vonNeumann1932} if we take the averaging over modular times to act on the Hilbert space through $\frac{1}{2\Lambda}\int_{-\Lambda}^{\Lambda}ds\, \Delta^{is}$. The corresponding limit would be a projector onto the $\Delta^{is}$-invariant subspace. For a discussion of the physical importance of distinguishing the strong and the weak operator topology in the context of modular zero modes we refer to \cite{Lashkari:2019ixo}.} (see footnote \ref{footnote:topologies} for a definition). The convergence of this integral is thus understood within expectation values, in the sense that the operator $\mathcal{E}_0(a)$ is determined by the condition that
\beq \label{eq:integralconvergence}
\psi(\mathcal{E}_{0}(a)b) = \lim_{\Lambda\to\infty}\frac{1}{2\Lambda}\int_{-\Lambda}^{\Lambda}ds\,\psi(\sigma_s(a)b)~, \qquad \forall b\in \mathcal{A}~.
\eeq
It is now straightforward to check that the map $\mathcal{E}_0$ satisfies the right properties. By definition, it leaves the centralizer invariant. Indeed, given a modular zero mode $a\in \mathcal{A}^{\psi}$ it follows that 
\beq 
\mathcal{E}_{0}(a)=\lim_{\Lambda \to \infty, \mathrm{weak}}\frac{1}{2\Lambda}\int_{-\Lambda}^{\Lambda}ds\,\sigma_s(a)= a \lim_{\Lambda \to \infty, \mathrm{weak}}\frac{1}{2\Lambda}\int_{-\Lambda}^{\Lambda}ds\,1 = a~.
\eeq
In addition, since $\mathds{1}$ is always in the centralizer, we automatically have $\mathcal{E}_0(\mathds{1})=\mathds{1}$. To show the bimodule property \eqref{eq:bimoduleproperty}, we consider $a_1,a_2\in \mathcal{A}^{\psi}$ and $b\in \mathcal{A}$, and observe that    
\begin{align}
\mathcal{E}_{0}(a_1ba_2)&=\lim_{\Lambda \to \infty, \mathrm{weak}}\frac{1}{2\Lambda}\int_{-\Lambda}^{\Lambda}ds\,\sigma_s(a_1)\sigma_s(b)\sigma_{s}(a_2) \nonumber \\
&=a_1\left[\lim_{\Lambda \to \infty, \mathrm{weak}}\frac{1}{2\Lambda}\int_{-\Lambda}^{\Lambda}ds\,\sigma_s(b)\right]a_2=a_1\mathcal{E}_0(b)a_2~.
\end{align}
Furthermore, using the invariance of the state under modular flow, $\psi\circ \sigma_s =\psi$, we derive the identity
\beq 
\psi\circ \mathcal{E}_0(a)=\lim_{\Lambda \to \infty}\frac{1}{2\Lambda}\int_{-\Lambda}^{\Lambda}ds\,\psi(\sigma_s(a))=\psi(a)\lim_{\Lambda \to \infty}\frac{1}{2\Lambda}\int_{-\Lambda}^{\Lambda}ds\,1 =\psi(a)~,
\eeq
for all $a\in \mathcal{A}$ which proves \eqref{eq:psiE0}. Hence, \eqref{eq:zeromodeprojector} satisfies all the required properties. 

Let us now come back to the modular Fourier decomposition as defined in \eqref{eq:eigenvalue}. In terms of the modular Fourier modes $a_{\omega} \in \mathcal{A}_{\omega}$, the conditional expectation acts by setting $\omega=0$. We have 
\beq 
\mathcal{E}_{0}(a_{\omega})=\lim_{\Lambda \to \infty}\frac{1}{2\Lambda}\int_{-\Lambda}^{\Lambda}ds\,\sigma_s(a_{\omega}) = a_{\omega} \lim_{\Lambda \to \infty}\frac{1}{2\Lambda}\int_{-\Lambda}^{\Lambda}ds\,e^{i\omega s}=\begin{cases} a_{\omega} \hspace{10pt} &\mathrm{if} \hspace{10pt}  \omega = 0~,  \\
0 \hspace{10pt}  &\mathrm{if} \hspace{10pt}  \omega \neq 0~.\end{cases}~.
\eeq
This shows that the zero mode projector is well-defined on the space of modular Fourier modes. To compare with the usual way of expressing the modular zero mode projector, we consider the following spectral decomposition of the modular operator:
\beq 
\Delta_{\psi} = \int d\Pi(\omega)\, e^{-\omega}~,
\eeq
where $d\Pi(\omega)$ is a projection-valued measure, and the domain of the integral covers the spectrum of the modular Hamiltonian. We can now express the conditional expectation in terms of
\beq 
\mathcal{E}_{0}(a) = \int \int d\Pi(\omega)\,a\,d\Pi(\omega') \lim_{\Lambda \to \infty}\frac{1}{2\Lambda}\int_{-\Lambda}^{\Lambda}ds\, e^{i(\omega'-\omega)s}~,
\eeq
and evaluate the integral over modular time to be given by
\beq \label{eq:intergraldelta}
\lim_{\Lambda \to \infty}\frac{1}{2\Lambda}\int_{-\Lambda}^{\Lambda}ds\, e^{i(\omega'-\omega)s} = \begin{cases} 1 \hspace{10pt} \mathrm{if} \hspace{10pt}  \omega = \omega'~,  \\
0 \hspace{10pt}  \mathrm{if} \hspace{10pt}  \omega \neq \omega'~.\end{cases}
\eeq
Heuristically, we can think of the above equation in terms of a Dirac $\delta$-function that is renormalized with an infinite constant  $\delta(0)^{-1}\delta(\omega-\omega')$. After this somewhat formal manipulation, the $\delta$-function removes one of the frequency integrals and we arrive at the following form of the projection operator:
\beq \label{eq:condPVM}
\mathcal{E}_{0}(a) = \delta(0)^{-1}\int d\Pi(\omega)\,a\,d\Pi(\omega)~,
\eeq
which is the formula used in \cite{Czech:2023zmq} to implement the modular zero mode projector. To match the expression there, one writes the projection-valued measure $d\Pi(\omega) = d\omega \, |\omega\rangle \langle \omega|$ in terms of eigenstates $|\omega\rangle$ of the modular Hamiltonian, which exist in the case that the underlying von Neumann algebra is type I. This leads to the expression:
\beq \label{eq:eigenstates}
\mathcal{E}_{0}(a) = \delta(0)^{-1}\int d\omega \langle \omega | a |\omega \rangle |\omega\rangle \langle \omega |~.
\eeq
Let us stress, however, that the projection map onto the space of modular zero modes can be defined more abstractly in terms of the conditional expectation \eqref{eq:condexp0}, and its definition does not rely on any specific assumptions about the algebra type. We have thus established that the modular zero mode projector make sense, even when we cannot naively use the form of the projection operator in \eqref{eq:eigenstates}. In that sense, the present paper provides a natural generalization of the earlier discussions.

\subsection{The modular Berry curvature} \label{sec:modularBerry}

We are now ready to introduce the modular Berry curvature. To this end, we excite the state $|\psi\rangle$ by acting on it with some operator:
\beq
|\psi\rangle \to |u\rangle \equiv u|\psi\rangle~,
\eeq
where $u$ is a unitary in the algebra $\mathcal{A}$. It is easy to check that the state $|u\rangle$ is cyclic and separating provided that $|\psi\rangle$ is. To explore the space of states, we assume that the unitary $u=u(\lambda)$ depends on some external parameter $\lambda$. It turns out that the modular operator associated to the excited state is related to the original modular operator by conjugation with the unitary $u$. To see this, one first computes the Tomita operator according to \eqref{eq:Spsi}. It is given by the conjugated version of the original map,
\beq \label{eq:uSu}
S_{\psi}(\lambda)\equiv u(\lambda)S_{\psi}u(\lambda)^{\dagger}~,
\eeq
as can be seen from verifying the definition:
\beq 
S_{\psi}(\lambda) a|u(\lambda)\rangle = u(\lambda)S_{\psi} u(\lambda)^{\dagger}a u(\lambda)|\psi\rangle = u(\lambda)u(\lambda)^{\dagger}a^{\dagger}u(\lambda)|\psi\rangle = a^{\dagger}|u(\lambda)\rangle~.
\eeq
The corresponding modular operator $\Delta_{\psi}(\lambda)$ is now computed to be
\beq \label{eq:uDeltau}
\Delta_{\psi}(\lambda) =S_{\psi}(\lambda)^{\dagger}S_{\psi}(\lambda)= u(\lambda)S_{\psi}^{\dagger}S_{\psi}u(\lambda)^{\dagger} = u(\lambda) \Delta_{\psi} u(\lambda)^{\dagger}~,
\eeq
where we have used the unitarity of the perturbations. Similarly, by taking the logarithm on both sides it follows that the modular Hamiltonian associated to the perturbed state is given by 
\beq 
K_{\psi}(\lambda)=u(\lambda)K_{\psi}u(\lambda)^{\dagger}~.
\eeq 
Importantly, not all excitations lead to distinct modular operators, and in particular $\Delta_{\psi}$ is left invariant when the unitary is a modular zero mode $u\in \mathcal{A}^{\psi}$. As a consequence, for each value of the external parameter $\lambda$ there is a choice of modular zero mode frame that can be encoded in a suitable gauge field. The gauge field defines a one-form on the space of perturbations, so to define it we consider small perturbations around a fixed modular operator:
\beq 
u(\epsilon) = e^{i\epsilon X}=\mathds{1}+i\epsilon X+\mathcal{O}(\epsilon^2)~,
\eeq
where $X$ is a Hermitian generator and $\epsilon$ a small parameter. When the unitary is a modular zero mode, its generator commutes with the modular Hamiltonian. On the level of the generators, we can define a zero mode projection $P_0$ by taking $P_0(X)$ to be the generator of the zero mode piece of the unitary $u(\epsilon)$ obtained through the conditional expectation $\mathcal{E}_0$:
\beq \label{eq:P0definition}
P_0(X)\equiv\frac{d}{d\epsilon}\Big|_{\epsilon = 0}\mathcal{E}_0(u(\epsilon))~.
\eeq
It follows from the definition that $P_0(X)$ commutes with the modular Hamiltonian $K_{\psi}$. In practice, one mostly uses the form of the projection operator in \eqref{eq:zeromodeprojector} provided that the resulting expression is well-defined.

As a simple example, let us write down an explicit expression for the projector when the modular zero mode algebra is trivial, i.e., when $\mathcal{A}^{\psi}=\mathbb{C}\cdot \mathds{1}$. In that case, the conditional expectation maps onto an operator proportional to the identity: $\mathcal{E}_0=\alpha \mathds{1}$. In order to satisfy \eqref{eq:psiE0}, the coefficient $\alpha$ must be given by the expectation value with respect to the state $\psi$. It then follows that 
\beq \label{eq:trivialP0}
P_0(X)=\langle \psi|X|\psi\rangle \mathds{1}~, 
\eeq
which is a formula that we will use later on. 

Let us come back to the modular zero mode ambiguity. As mentioned before, at each point $\Delta(\lambda)$ along the curve of modular operators \eqref{eq:uDeltau} there is a choice of zero mode frame, that is not visible in the base space. We introduce the connection
\beq 
\Gamma \equiv P_0(u(\lambda)^{\dagger}\delta u(\lambda))~,
\eeq
that keeps track of the frame ambiguity at each point. 
Given a change of zero mode frame $u(\lambda)\mapsto u(\lambda)v(\lambda)$ in terms of some explicit zero mode $v(\lambda) \in \mathcal{A}^{\psi}$, $\Gamma$ transforms as
\beq 
\Gamma \mapsto v(\lambda)^{\dagger}\Gamma v(\lambda)+v(\lambda)^{\dagger}\delta v(\lambda)~.
\eeq
The condition of parallel transport now states that  $\tilde{u}^{\dagger}\delta \tilde{u}$ cannot pick up an additional zero mode piece when transported along a curve $\tilde{u}$ \cite{Czech:2019vih}, i.e., 
\beq 
P_{0}(\tilde{u}^{\dagger}\delta \tilde{u})=0~.
\eeq
To compute the associated curvature, we need to consider a two-parameter family of perturbations specified by two unitaries $u(\lambda_1)$ and $u(\lambda_2)$. If we denote their generators by $X_1, X_2$, after subtracting the zero mode piece such that $P_0(X_1)=P_0(X_2)=0$, the modular Berry curvature is given by the formula
\beq \label{eq:modularBerrycurvature1}
F = P_0([X_1,X_2])~.
\eeq
Note that the Berry curvature involves a commutator of the two perturbations $X_1,X_2$. In the specific case that we only have trivial zero modes, $\mathcal{A}^{\psi}=\mathbb{C}\cdot \mathds{1}$, it is therefore not necessary to impose the condition $P_0(X_1)=P_0(X_2)=0$ when using equation \eqref{eq:modularBerrycurvature1}. The zero mode pieces $P_0(X_1), P_0(X_2)$, which are both proportional to the identity operator by virtue of \eqref{eq:trivialP0}, drop out automatically. 

An important matrix element of the operator $F$ is its expectation value in the original state $|\psi\rangle$. We define 
\beq 
F_{\psi} \equiv \langle \psi| F |\psi \rangle ~.
\eeq
Note that the expectation value in the state $|\psi\rangle$ is compatible with the zero mode projection in the sense of equation \eqref{eq:psiE0}. Hence, the non-zero mode pieces drop out automatically when taking the expectation value, and we can simply write
\beq \label{eq:modularBerrycurvature2}
F_{\psi} = \langle \psi|[X_1,X_2]| \psi \rangle ~.
\eeq
This is a convenient way of representing the modular Berry phase, and one that we will mostly use. We would like to stress that the operator $F$ contains more information than $F_{\psi}$. An example where this occurs will be discussed in Section \ref{sec:modchaos}. 

It is interesting to note that the modular Berry phase can be used to probe the structure of the underlying von Neumann algebra. In particular, if one takes the generators in \eqref{eq:modularBerrycurvature2} to be bounded operators $a_1,a_2\in \mathcal{A}$, the corresponding Berry phase $F_{\psi}$ can be used to determine the type of the algebra. To be precise, if there exists a state $|\phi\rangle$ such that the Berry phase vanishes for all algebra elements the corresponding algebra must be type I or type II\footnote{This does not mean that the Berry phase has to vanish for all states. For example, one can consider a spin in an external magnetic field. While most states exhibit non-zero Berry phases, there does exist a state of the system, namely the purification of the maximally mixed state, that has vanishing Berry phase for all Pauli matrices.}. The reason is that the vanishing modular Berry phase condition
\beq 
F_{\phi}=\langle \phi| [a_1,a_2] |\phi\rangle = 0~, \qquad \forall a_1,a_2\in \mathcal{A}~, 
\eeq
is equivalent to saying that the expectation value $a\mapsto \phi(a)$ with respect to the state $|\phi\rangle$ defines a trace on the algebra $\mathcal{A}$. In fact, because we are working with a state\footnote{Although in the present paper we have worked with states, one can imagine a generalization of the modular Berry phase $F_{\phi}$ to arbitrary weights $\phi$. Given a strictly semi-finite weight with vanishing Berry curvature, there is no need for the resulting trace to be finite, so one would instead conclude that the algebra is type I or type II (including the possibility of type I$_{\infty}$ or type II$_{\infty}$).}, the trace induced on the centralizer is normalized to one, so the algebra is type I$_n$ with $n<\infty$ or type II$_1$. We thus conclude:
\begin{quotation}
\noindent \textit{If there exists a state such that the modular Berry phase vanishes for all perturbations in the algebra, the algebra must be type I$_n$ or type II$_1$.}
\end{quotation}
In particular, a similar diagnostic for distinguishing algebra types based on \eqref{eq:modularBerrycurvature2} was proposed in \cite{Banerjee:2023eew}. We have identified the relevant geometric quantity to be the \emph{modular} Berry curvature, which we define for arbitrary von Neumann algebras and show that it reduces to \eqref{eq:modularBerrycurvature2} upon taking expectation values. While studying the vanishing of the Berry phase gives an in-principle way of determining the algebra type, it might not be of the most practical use, since one needs to know the Berry curvature for all possible deformations.

\subsection{A quantum information metric}

We would like to end this section by arguing that the modular Berry phase provides a natural metric on the space of quantum states. The idea is to take one of the perturbations in \eqref{eq:modularBerrycurvature2} to be of the form $[X_1,K_{\psi}]$ where $K_{\psi}$ is the modular Hamiltonian. This leads to the following quantity
\beq 
G_{\psi} = \langle \psi | [[X_1,K_{\psi}],X_2]| \psi \rangle~,
\eeq
which was introduced in \cite{Czech:2023zmq}. It is easy to see that this expression is symmetric under interchanging $1 \leftrightarrow 2$. Indeed, by writing out the commutator and using that the modular Hamiltonian annihilates the state via  $K_{\psi}|\psi\rangle =0$, it follows immediately that 
\begin{align} 
G_{\psi}& = \langle \psi | \left(-K_{\psi}X_1X_2 + X_1K_{\psi}X_2 +X_2K_{\psi}X_1 - X_2X_1K_{\psi}\right)
| \psi \rangle \nonumber \\ 
&=\langle \psi |\left(X_1K_{\psi}X_2+X_2K_{\psi}X_1\right)| \psi \rangle~. \label{eq:Gderivation}
\end{align}
This last expression is manifestly symmetric under exchanging $X_1 \leftrightarrow X_2$. 

It turns out that the above assignment provides a notion of distance on the space of quantum states. To see this, we consider excited state of the form $|u\rangle = u |\psi\rangle$ as before, where $u$ is some unitary in the algebra.
We now introduce what is known as the \emph{relative modular operator} $\Delta_{\phi|\psi}$ associated to two states $|\phi\rangle$ and $|\psi\rangle$. It is a generalization of the modular operator in the sense that we get it back when we set the two states to be equal, i.e., $\Delta_{\psi}=\Delta_{\psi|\psi}$. We are interested in the case where one of the states is given by the excited state $|u\rangle$ with respect to $|\psi\rangle$. To define the relative modular operator in this case, we first have to give an expression for the \emph{relative Tomita map} $S_{u|\psi}$. It is defined by the condition
\beq 
S_{u|\psi}a|\psi\rangle = a^{\dagger}|u\rangle~,
\eeq
for all $a\in \mathcal{A}$. Given the specific form of the state $|u\rangle$, a short computation shows that the map $S_{u |\psi} = S_{\psi}u^{\dagger}$ satisfies
\beq 
S_{u|\psi} a|\psi\rangle = S_{\psi}u^{\dagger}a|\psi\rangle = a^{\dagger} u|\psi \rangle = a^{\dagger} |u\rangle~,
\eeq
as required. Similar to the usual modular operator, its relative version is defined by
\beq 
\Delta_{u|\psi}=S_{u|\psi}^{\dagger}S_{u|\psi}=uS_{\psi}^{\dagger}S_{\psi}u^{\dagger}= u \Delta_{\psi} u^{\dagger}~,
\eeq
so it is simply obtained by conjugation with $u$. By taking the logarithm on both sides, one finds that the relative modular Hamiltonian is given by 
\beq \label{eq:Ku|psi}
K_{u|\psi}=-\log \Delta_{u|\psi}= uK_{\psi} u^{\dagger}~.
\eeq
The \emph{relative entropy} can be expressed in terms of the relative modular operator via the formula 
\beq 
S(u|\psi)\equiv -\langle \psi|\log \Delta_{u|\psi}|\psi\rangle = \langle \psi|K_{u|\psi}|\psi\rangle~.
\eeq 
Assuming that the unitary generates a small perturbation of the original state in the sense that $u=\mathds{1}+i\varepsilon X +\cdots$, where $\varepsilon$ is a small parameter, we can expand the relative modular Hamiltonian in \eqref{eq:Ku|psi} to second order in $\varepsilon$. One finds that
\beq 
K_{u|\psi} = K_{\psi}+i\varepsilon[K_{\psi},X]-\frac{\varepsilon^2}{2}\left(X^2K_{\psi}- 2 X K_{\psi}X+ K_{\psi}X^2\right)+\cdots~.
\eeq
Plugging this result into the formula for the relative entropy, gives the following expression:
\begin{align}
S(u|\psi) & = \langle \psi|K_{\psi}|\psi\rangle+i\varepsilon\langle \psi|[K_{\psi},X]|\psi\rangle-\frac{\varepsilon^2}{2}\langle \psi|\left(X^2K_{\psi}- 2XK_{\psi}X+ K_{\psi}X^2\right)|\psi\rangle+\cdots \nonumber \\
& = \varepsilon^2\langle \psi|XK_{\psi}X|\psi\rangle+\cdots~,
\end{align}
where we have used the condition  $K_{\psi}|\psi\rangle = 0$ to argue that most terms in the above expression vanish, except for the second term at second order in the parameter $\varepsilon$. From the second order variation of the relative entropy, which by the previous computation takes the simple form
\beq \label{eq:2ndorder}
\frac{d^2}{d\varepsilon^2}\Big|_{\varepsilon=0} S(u|\psi) = \langle \psi|XK_{\psi}X|\psi\rangle~,
\eeq
one can construct a symmetric quantity that is known as the \emph{quantum Fisher information metric}. It is given by 
\beq \label{eq:QFmetric}
g_{\mathrm{QF}}(X_1,X_2)\equiv \langle \psi|\left(X_1K_{\psi}X_2+X_2 K_{\psi}X_1\right)|\psi\rangle~,
\eeq
and defines a metric on the space of quantum states. It is the quantum analogue of the usual Fisher information metric on the classical information geometry. Comparing with \eqref{eq:Gderivation}, we find that the expression in \eqref{eq:QFmetric} agrees with the symmetric quantity that we derived from the Berry curvature, i.e.,
\beq 
G_{\psi}(X_1,X_2)= g_{\mathrm{QF}}(X_1,X_2)~.
\eeq
This is a simple proof that the modular Berry curvature is related to the quantum Fisher information metric if we take one of the perturbations to be of the form $[X,K_{\psi}]$. This result was already obtained in \cite{Czech:2023zmq} following a slightly different route, but the above derivation in terms of the relative modular operator seems to be more direct. The above relation between the modular Berry curvature and a metric on the space of quantum states points towards an interesting link with complexity that we hope to explore in future work.

%% file: GFF.tex
\section{An emergent symplectic form in the large $N$ limit}
\label{sec:symform}
One important aspect of the recent algebraic approach to holography is the emergence of a type III$_1$ boundary algebra in the large $N$ limit. As a prototypical example, we consider two copies of a boundary CFT in the thermofield double state above the Hawking-Page temperature. The bulk dual via the holographic dictionary is described by the eternal black hole geometry. It was argued in \cite{Leutheusser:2021frk} that the algebra of single-trace operators acting on either side of the thermofield double state has the structure of a type III$_1$ von Neumann algebra in the strict $N=\infty$ limit. The type III$_1$ nature of the algebra and the existence of a half-sided modular inclusion in the boundary theory (which will be discussed in more detail in Section \ref{sec:modchaos}) are closely related to the horizon structure in the emergent bulk spacetime.

In the present section, we would like to show that the modular Berry curvature, when computed for the type III$_1$ boundary algebra, encodes information about the bulk geometry, in the sense that the parallel transport of modular operators in the boundary gives rise to an emergent symplectic form in the large $N$ limit. To establish this, we first give a precise definition of the large $N$ algebra acting on the boundary CFT, and compute the modular operator associated to a subclass of excited states that are obtained from the global state by acting with a unitary in the algebra. We then compute the corresponding modular Berry curvature. Using the extrapolate dictionary, we show that it agrees with a suitably defined bulk symplectic form. While our results are phrased in the context of the eternal black hole spacetime, the modular Berry phase can be computed in more general setups where one has an emergent type III$_1$ algebra. It would be interesting, for example, to use the above technology to study general time band algebras in the CFT \cite{Leutheusser:2021frk,Leutheusser:2022bgi,Gesteau:2024rpt}.

\subsection{The large $N$ von Neumann algebra}
\label{sec:largeN}

We consider two copies of a CFT, denoted by CFT$_L$ and CFT$_R$, in the thermofield double state $|\Psi_{\beta}\rangle$ at inverse temperature $\beta$. At finite $N$, the Hilbert space of the total system can be factorized as $\mathcal{H}=\mathcal{H}_L\otimes \mathcal{H}_R$ and the thermofield double state takes the familiar form
\beq \label{eq:TFD}
|\Psi_{\beta}\rangle = \frac{1}{\sqrt{Z}}\sum_{n}e^{-\beta E_n/2}|n\rangle_{L}|n\rangle_{R}~.
\eeq
The single-trace operators that act on the left CFT generate an algebra $\mathcal{A}_{L}$. Similarly, one can define an algebra of single-trace operators $\mathcal{A}_R$ acting on the right CFT. In the $N\to \infty$ limit, we do not have a simple tensor product of the left and right Hilbert spaces, but the thermofield double state \eqref{eq:TFD} has a well-defined limit that we still denote by $|\Psi_{\beta}\rangle$. We can now construct the Hilbert space in terms of excitations on top of the thermofield double state by acting with single-trace operators in one of the boundary algebras. Mathematically, this corresponds to the so-called the Gelfand-Naimark-Segal (GNS) construction. The idea is to define a linear space consisting of states
\beq 
|a\rangle \equiv a|\Psi_{\beta}\rangle~,
\eeq
and an inner product by taking the expectation value with respect to $|\Psi_{\beta}\rangle$:
\beq \label{eq:GSN_innerproduct}
\langle a_1|a_2\rangle \equiv \langle \Psi_{\beta}|a_1^{\dagger} a_2|\Psi_{\beta} \rangle~.
\eeq
Provided that the state $|\Psi_{\beta}\rangle$ is separating, the above assignment defines an inner product. (The GNS construction instructs that whenever the inner product defined above is degenerate, i.e., has null states, one first needs to divide out the subspace of null states.) We then take the completion with respect to the inner product by including all limits. Let us denote the resulting GNS Hilbert space by $\mathcal{H}^{\rm GNS}$. The algebra $\mathcal{A}_L$ associated to the left CFT now acts via left multiplication, while the operators on the right CFT are represented by right multiplication:
\beq 
a_1|a_2\rangle = |a_1 a_2\rangle~,\qquad a_1\in \mathcal{A}_L~, \qquad a_1'|a_2\rangle = |a_2a_1'\rangle~,\qquad a_1'\in \mathcal{A}_R~.
\eeq
It then follows that the left and right algebras are commutants of each other, $\mathcal{A}_L'=\mathcal{A}_R$. To leading order in $1/N$, the corresponding boundary theory is described by a \emph{generalized free field} (GFF) theory \cite{Greenberg:1961mr, Aharony:1999ti, Duetsch:2002hc}. This means that the theory is completely fixed in terms of the two-point functions, while all higher point correlation functions, which are suppressed in powers of $N$, vanish.

It is important to note that not all operators survive in the $N\to \infty$ limit. If the operator is sufficiently complicated, meaning that the number of single-trace operators to build it scales with $N$, it may not be well-defined in the large $N$ von Neumann algebra. The most prominent example involves the density matrix of the thermofield double state. At finite $N$, it can be defined by tracing over the complementary subsystem to obtain
\beq 
\rho_{L}=\mathrm{tr}_{R} |\Psi_{\beta}\rangle \langle \Psi_{\beta}|= \frac{1}{Z}\sum_{n}e^{-\beta E_n}|n\rangle_L \langle n|_{L}~,
\eeq
but this does not define an operator in the large $N$ algebra. As mentioned at the start of Section \ref{sec:algebraBerry}, the existence of density matrices is related to the underlying von Neumann algebra being type I or type II. Instead, it is useful to consider the two-sided modular operator $\Delta_{\Psi}$ associated to the state $|\Psi_{\beta}\rangle$, which can be defined for the large $N$ algebra and will be studied momentarily. 

Before moving on to the Berry curvature computation, let us be a bit more precise about the definition of the boundary operator algebra and discuss which operators to include. As usual, the local operators $\mathcal{O}(x)$ themselves are somewhat ill-behaved, and one needs to smear them with respect to suitable smearing functions. We therefore consider operators of the form
\beq  \label{eq:smearing}
\Phi(\lambda)\equiv\int dx\,\lambda(x)\mathcal{O}(x)~,
\eeq
which are more natural objects to consider from a QFT point of view. The precise form of the algebra depends on the kind of smearing functions $\lambda(x)$ that we consider. A common choice of smearing functions is given by the $*$-algebra $\mathcal{S}$ of complex \emph{Schwartz functions}. These are smooth bounded functions that decay faster than any power law at infinity in any non-compact directions (see, for example, Section 3.2 of \cite{deBoer:2021zlm} for more details). A useful property of Schwartz functions is that their Fourier transform is again a Schwartz function. Note that the smearing functions are functions of both space and time, so in order for the causal propagator, that we will define momentarily, to give rise to a symplectic form, we have to additionally introduce equivalence classes of
smearing functions by modding out the functions that are in the range of the equations of
motion. This ensures that the resulting two-form is non-degenerate. The smeared operators in \eqref{eq:smearing} are generally still unbounded, but one can now formally define a $*$-algebra by considering sums of products of $\Phi(\lambda)$ using the algebraic structure of the smearing functions. In particular, assuming that $\mathcal{O}$ is a Hermitian operator we have 
\beq 
\Phi(\lambda)^{\dagger}= \Phi(\lambda^*)~.
\eeq

To introduce the structure of a von Neumann algebra, one needs to take the completion within a suitable Hilbert space on which the action of the operators can be represented. This will be the GNS algebra mentioned above. In the case where we are dealing with a GFF theory, the corresponding GNS Hilbert space takes the form of a Fock space of multi-particle states. The Hilbert space is generated by linear combinations of states of the form 
\beq \label{eq:multi-particle}
|\lambda_1,\cdots, \lambda_n\rangle \equiv \Phi_+(\lambda_1)\cdots \Phi_+(\lambda_n)|\Psi_{\beta}\rangle~, 
\eeq
where $\Phi_{+}(\lambda)$ is the part of $\Phi(\lambda)$ containing creation operators only. Note that the since the creation operators commute amongst themselves the above states live in the symmetric tensor product of the single particle Hilbert space. The inner product on the Fock space is determined by the two-point function 
\beq \label{eq:hermitianmetric}
\langle \lambda_1| \lambda_2 \rangle \equiv \langle \Psi_{\beta} | \Phi_-(\lambda_1^*)\Phi_+(\lambda_2) |\Psi_{\beta} \rangle = \langle \Psi_{\beta} |\Phi(\lambda_1^*)\Phi(\lambda_2)|\Psi_{\beta} \rangle ~,
\eeq
as follows from \eqref{eq:GSN_innerproduct}. After dividing out possible null states to make \eqref{eq:hermitianmetric} into a non-degenerate inner product, we can express the inner product between more general multi-particle states \eqref{eq:multi-particle} as 
\beq \label{eq:innerproduct}
\langle\lambda_1',\lambda_2',\ldots,\lambda_n'|\lambda_1,\lambda_2,\ldots,\lambda_n\rangle = \prod_{i=1}^n \langle \lambda_i'|\lambda_i\rangle~.
\eeq 
Taking the completion with respect to this inner product defines the Fock space $\mathcal{H}^{\rm GNS}$ of multi-particle states. Note that the factorized form of the inner product in \eqref{eq:innerproduct} is only correct to leading order in $1/N$. 

To construct the von Neumann algebra of operators associated to the left CFT, we start from the smeared operator \eqref{eq:smearing} and exponentiate it to get a bounded operator. The resulting unitaries (provided that the source $\lambda$ is real)
\beq 
u(\lambda)= e^{i\Phi(\lambda)}~,
\eeq
naturally act on the GNS Hilbert space, so they generate a subalgebra $\mathcal{U}$ of $B(\mathcal{H}^{\rm GNS})$. By taking the weak closure (see footnote \ref{footnote:topologies} for a definition) of the algebra $\mathcal{U}$ within the $\mathcal{H}^{\rm GNS}$ we obtain the large $N$ von Neumann algebra. Equivalently, one can express $\mathcal{A}_L=\mathcal{U}''$ by the double commutant theorem \cite{Neumann1930}.
  
\subsection{Computing the modular Berry phase}

Let us now set up a nontrivial parallel transport. We consider the following class of states:
\beq \label{eq:statedeformation}
|\delta\lambda \rangle \equiv u(\delta\lambda)|\Psi_{\beta}\rangle~,
\eeq 
which are obtained by acting on the TFD state with a unitary $u(\delta\lambda)\in \mathcal{A}_L$. As explained before in \eqref{eq:uDeltau}, the modular operator associated with the excited state $|\delta\lambda\rangle$ can be obtained from the original modular operator $\Delta_{\Psi}$ by conjugating with the unitary,
\beq \label{eq:udelta}
\Delta(\delta\lambda)  = u(\delta\lambda) \Delta_{\Psi} u(\delta\lambda)^{\dagger}~.
\eeq
Expanding the unitaries in the parameter $\delta \lambda$, one can now express the perturbed modular Hamiltonian as a series of nested commutator with the original modular Hamiltonian $K_{\Psi}$:
\beq
K(\delta\lambda)=K_{\Psi}+i[\Phi(\delta\lambda),K_{\Psi}]-\frac{1}{2}[\Phi(\delta\lambda),[\Phi(\delta\lambda),K_{\Psi}]]+\cdots.
\eeq
Assuming that $\delta \lambda$ is infinitesimal, we keep only the first order perturbation and write the change of the modular Hamiltonian as 
\beq 
\delta K= i[\Phi(\delta\lambda),K_{\Psi}]~.
\eeq
This is the infinitesimal version of the deformation in \eqref{eq:udelta}. We therefore identify the generator that implements the change in the modular Hamiltonian by
\beq \label{eq:paralleltransport}
X = i \Phi(\delta \lambda)~.
\eeq
In addition, we need to impose the condition $P_0(X)=0$ when there are non-trivial modular zero modes.

To probe the Berry curvature, we now consider two different perturbations $\delta_1\lambda$ and $\delta_2\lambda$, and denote the corresponding generators by $X_1=i\Phi(\delta_1\lambda)$ and $X_2=i\Phi(\delta_2\lambda)$, respectively. In order to evaluate \eqref{eq:modularBerrycurvature1}, we state an important property of the large $N$ algebra, namely that the commutator of two fundamental fields (which in the present context are the local operators $\mathcal{O}(x)$ that are used to generate the algebra) is a c-number. We have 
\beq \label{eq:comm}
[\mathcal{O}(x),\mathcal{O}(y)]=-i\propegator(x-y)\mathds{1}~,
\eeq
where $\propegator(x-y)$ denotes the causal propagator, which depends on the details of the theory. Note that it is antisymmetric when $x\leftrightarrow y$. We can use \eqref{eq:comm} to calculate the commutator as
\beq 
[X_1,X_2]=-[\Phi(\delta_1\lambda),\Phi(\delta_2\lambda)]=i\int dx\int dy\,\delta_1\lambda(x)\propegator(x-y)\delta_2\lambda(y)\,\mathds{1}~.
\eeq
Note that the above expression is proportional to the identity operator $\mathds{1}$, and by definition the identity operator is always a zero mode:
\beq 
\mathcal{E}_0(\mathds{1})= \mathds{1}~.
\eeq
This shows that in the present case the modular zero mode projection acts trivially. We conclude that the modular Berry curvature associated to two coherent state deformations of the form \eqref{eq:statedeformation} is given by 
\beq \label{eq:resultF}
F = P_0([X_1,X_2])=i\int dx\int dy\,\delta_1\lambda(x)\propegator(x-y)\delta_2\lambda(y)\,\mathds{1}~.
\eeq
  
Let us end this section with a two remarks. First, note that the result in \eqref{eq:resultF} is quite special in the sense that the Berry curvature is in the center of the algebra, which means that the operator is shared between the left and the right system, i.e.,
\beq 
F\in \mathbb{C}\cdot \mathds{1}=\mathcal{A}_L\cap \mathcal{A}_R~.
\eeq
This is a somewhat special situation, because the computation is carried out directly in the large $N$ von Neumann algebra through the assumption \eqref{eq:comm}. Including $1/N$ corrections modifies the commutator in \eqref{eq:comm}, and the appearance of additional modular zero modes in this expression will lead to a different result for the Berry curvature. We expect that the correct language to describe this structure algebraically is through the crossed product algebra \cite{Witten:2021unn}. We hope to address this in future work. 

Secondly, it is often useful to consider a specific matrix element of the operator $F$ that is obtained by evaluating it in the original state $|\Psi_{\beta}\rangle$. It is interesting to point out that in the present case, as the operator is diagonal, one does not lose any information by doing this. It simply removes the identity operator in the process:
\beq \label{eq:FOmega}
F_{\Psi} \equiv \langle \Psi_{\beta} | F |\Psi_\beta \rangle = i \int dx\int dy\,\delta_1\lambda(x)\propegator(x-y)\delta_2\lambda(y)~.
\eeq
It follows from the properties of the propagator $\propegator(x-y)$ that the above expression is antisymmetric under swapping the two perturbations $1\leftrightarrow 2$. Therefore, it can be used to define a symplectic structure on the space of perturbations.  

\subsection{The bulk description}
We will now discuss how the result of the above computation is related to the bulk symplectic form. Reflecting the GNS construction on the boundary, the bulk Hilbert space consists of small excitations on top of a fixed background geometry. Let us quickly review the details of the construction for the TFD state $|\Psi_{\beta}\rangle$. The dual bulk geometry is given by an eternal black hole in AdS$_{n+1}$, for which the metric can be written as 
\beq 
ds^2 = - f dt^2 + \frac{dr^2}{f}+r^2 d\Theta^2_{n-1}~,
\eeq
where $d\Theta^2_{n-1}$ denotes the metric for spatial manifold on the boundary, which we take to be a plane $\mathbb{R}^{n-1}$. We parametrize the boundary space of the CFT by a coordinate $x=(t,\vec{x})$, where $\vec{x}\in \mathbb{R}^{n-1}$, and denote the radial bulk direction by $r\in [0,\infty)$. Let us from now on denote the collective bulk coordinate by a capital letter, e.g., $X=(r,x)$.

To conform with our boundary description, we consider a bulk scalar operator $\phi$ that is dual to the single-trace operator $\mathcal{O}$. We will now consider the Fock space obtained from quantizing small perturbations on the background geometry. When restricted to the right region, the field operator $\phi$ can be expanded in modes using a formula of the form 
\beq \label{eq:bulkmodeexpansion}
\phi(X)=\int dq\int_0^{
\infty} \frac{d\omega}{2\pi}(\alpha_{\omega,q}(X)a_{\omega,q}+ \alpha_{\omega,q}^*(X)a_{\omega,q}^{\dagger})~,
\eeq
where the label $q$ denotes some additional quantum numbers coming from the extra spatial dimensions, and the oscillators satisfy the canonical commutation relations
\beq \label{eq:canonicalcom}
[a_{\omega,q},a_{\omega,q'}^{\dagger}]=2\pi \delta(\omega-\omega')\delta(q-q')~.
\eeq
The functions $\alpha_{\omega,q}(X)$ constitute a complete set of normalized mode functions on the right region.
There is a similar expression for the bulk fields in the left exterior region, with a different set of oscillators and mode functions. To construct the relevant Fock space, we use the Hartle-Hawking vacuum state in the bulk. To be precise, the bulk Hilbert space, denoted by $\mathcal{H}^{\rm Fock}$, is obtained by acting on the Hartle-Hawking state with the right oscillators. The von Neumann algebra generated by the operator $\phi$, and possibly other matter field operators if one decides to include those, in the right exterior region will be denoted by $\widetilde{\mathcal{A}}_R$. It describes excitations that do not significantly backreact onto the black hole geometry. Similarly, for the operators in the left region we will use the notation $\widetilde{\mathcal{A}}_L$. Locality implies that the resulting algebras are commutants, i.e., $\widetilde{\mathcal{A}}_R= \widetilde{\mathcal{A}}_L'$. 

\begin{figure}
    \centering
    \includegraphics[width=0.5\linewidth]{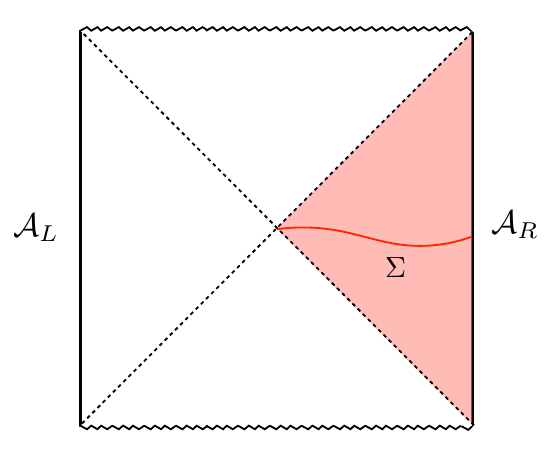}
    \caption{The eternal black hole geometry. The operators in the bulk algebra $\widetilde{\mathcal{A}}_R$ associated to the right exterior bulk region $D_R$ (depicted in red) can be reconstructed from the boundary algebra $\mathcal{A}_R$ of operators acting on the right CFT. We also identify a time slice $\Sigma$ for the right exterior region (red line) to which one can associate a symplectic form $\Omega$.}
    \label{fig:eternalBH}
\end{figure}

The general expectation is that the operator algebras constructed in this way are type III$_1$ von Neumann algebras. This is reflected in the local spacetime structure near the bifurcation surface of the eternal black hole. For example, there does not exist a well-defined notion of Schwarzschild vacuum state $|0\rangle_R\otimes |0\rangle_L$, which would require a factorization of the bulk Hilbert space across the left and right exterior region. Furthermore, one can show that the modular flow associated to the Hartle-Hawking state acts like a spacetime boost close to the bifurcation horizon. Because the geometric action of the boost is on both the left as well as on the right wedge region, the modular flow cannot be implemented using an operator in the left algebra. This shows that it is an outer autmorphism, which occurs whenever the von Neumann algebra is a type III factor.

Let us now connect to the boundary description in terms of the large $N$ von Neumann algebra constructed in Section \ref{sec:largeN}. Provided that the single-trace operator $\mathcal{O}$ is dual to the bulk operator $\phi$, one can identify the bulk and boundary Hilbert spaces of small excitations, $\mathcal{H}^{\rm Fock}=\mathcal{H}^{\rm GNS}$. In terms of operator algebras, the holographic duality states that the bulk and boundary algebras are equal:
\beq \label{eq:subalgreg}
\mathcal{A}_L = \widetilde{\mathcal{A}}_L~, \qquad \mathcal{A}_R = \widetilde{\mathcal{A}}_R~.
\eeq
In particular, this means that one can identify the oscillators in the generalized free field description of the boundary operator $\mathcal{O}$ with the oscillators in the bulk mode expansion \eqref{eq:bulkmodeexpansion}. We can therefore expand
\beq \label{eq:bdymodeexpansion}
\mathcal{O}(x)=\int dq \int_0^{
\infty} \frac{d\omega}{2\pi}(\gamma_{\omega,q}(x)a_{\omega,q}+ \gamma_{\omega,q}^*(x)a_{\omega,q}^{\dagger})~,
\eeq
where the oscillators $a_{\omega,q}, a_{\omega,q}^{\dagger}$ are identified with the ones in \eqref{eq:bulkmodeexpansion}, and the functions $\gamma_{\omega,q}(x)$ are a complete set of mode functions on the boundary manifold.
We stress that to make this identification one needs to represent the boundary operator $\mathcal{O}$ on the GNS Hilbert space $\mathcal{H}^{\rm GNS}$. Importantly, this representation is state-dependent; it depends, for example, on the inverse temperature $\beta$. The relation between $\mathcal{O}$ and $\phi$ is now given through the usual extrapolate dictionary:
\beq \label{eq:extrapolate}
\lim_{r\to \infty} r^{-\Delta_+}\phi(r,x) = \mathcal{O}(x)~, \qquad \lim_{r\to \infty} r^{-\Delta_+}\alpha_{\omega,q}(r,x) = \gamma_{\omega,q}(x)~,
\eeq
where $\Delta_+$ is the conformal dimension of the operator $\mathcal{O}$. Given that one has identified the boundary and bulk oscillators $a_{\omega,q}, a_{\omega,q}^{\dagger}$, one can interpret $\phi$ through its mode expansion as a boundary operator. This is a version of bulk reconstruction in the right region of the black hole exterior. We can see from the identification \eqref{eq:subalgreg} that the large $N$ von Neumann algebra is type III$_1$, provided that the bulk algebra is. This emergent type III$_1$ structure applies at leading order in the large $N$ limit, and $1/N$ corrections are expected to modify this picture. Note that in the above discussion we have considered the operator algebra associated to the full right boundary system. A similar statement can be made for more general subalgebras in the boundary system that are dual to geometric subregions in the bulk \cite{Leutheusser:2022bgi}. 

\subsubsection{The covariant phase space symplectic form}

We would now like to argue that the modular Berry curvature associated to the large $N$ von Neumann algebra can be represented geometrically in the bulk as a symplectic form. Before explaining the precise identification, let us quickly review the construction of the bulk covariant phase space symplectic form. A useful reference that we follow is \cite{Gieres:2021ekc}. To match our previous bulk discussion, we will consider a single real-valued scalar field $\varphi$ in the bulk, while noting that a straightforward generalization to more general field contents is possible. We assume that the dynamics are described by an action functional
\beq \label{eq:actionS}
S[\varphi]=\int dX \,\mathcal{L}(\varphi,\partial_{\mu}\varphi)~.
\eeq
If we introduce the canonical momentum degree of freedom
\beq 
\pi^{\mu} \equiv \frac{\delta \mathcal{L}}{\delta(\partial_{\mu}\varphi)}~,
\eeq
the phase space $\mathcal{M}$ consists of classical field configurations that are parametrized in terms of the canonical coordinates $\varphi^I=(\varphi,\pi^{\mu})$. The variation of the Lagrangian can be reorganized in terms of the Euler-Lagrange equation of motion $E$ plus a total derivative:
\beq 
\delta \mathcal{L} = E\delta \varphi + \partial_\mu j^{\mu}~.
\eeq
The term that appears as a total derivative is the symplectic potential $j^{\mu}\equiv \pi^{\mu} \delta \varphi$, which can be used to define a $\delta$-closed two-form on the classical phase space by setting
\beq \label{eq:current}
J^{\mu} \equiv -\delta j^{\mu} = \delta \varphi \wedge \delta \pi^{\mu}~.
\eeq
By definition $\delta J^{\mu}=0$, and it is also conserved on-shell, i.e.,
\beq 
\partial_{\mu}J^{\mu} =-\delta(\partial_{\mu}j^{\mu})=-\delta(\delta \mathcal{L})=0~.
\eeq
Given a codimension one spacelike surface $\Sigma$, we can define a symplectic two-form on phase space by integrating $J^{\mu}$ over $\Sigma$. We will take $D_R$ to be the right exterior region of the eternal black hole spacetime, and $\Sigma=\Sigma_R$ to be a Cauchy slice that extends from the boundary to the bifurcation surface (see Figure \ref{fig:eternalBH}). We define the covariant phase space symplectic form as
\beq \label{eq:covariantsymp}
\Omega \equiv  \int_{\Sigma} d\Sigma_{\mu} J^{\mu}~,
\eeq
where $d\Sigma_{\mu}$ is the volume form on $\Sigma$. To see that $\Omega$ indeed defines a symplectic form one observes that $\delta \Omega =0$, so $\Omega$ is $\delta$-closed, and $\Omega$ is non-degenerate in the absence of extra constraints. Note that the condition $\partial_{\mu}J^{\mu}=0$ implies that $\Omega$ is independent of the choice of Cauchy slice $\Sigma$.  Plugging in the definition of the current, one finds that it is given by
\beq \label{eq:covsymform}
\Omega = \int_{\Sigma} d\Sigma_{\mu}\left( \delta_1 \varphi \delta_2 \pi^{\mu} - \delta_2 \varphi \delta_1 \pi^{\mu}\right)~,
\eeq
which is the canonical way to represent the symplectic form in terms of conjugate variables $\varphi$ and $\pi^{\mu}$. We will comment on the boundary conditions for $\Omega$ that are set by the parallel transport equations at the end of Section \ref{sec:Berry = bulk}.

\subsubsection{The Poisson bracket}

Given a symplectic form, one can define a Poisson bracket on the space of functions. In the present case, the analogue of a function on phase space is played by a functional 
\beq \label{eq:functionalF}
\mathcal{F}:\mathcal{M}\to \mathbb{R}: \varphi \mapsto \mathcal{F}[\varphi]~,
\eeq
that maps a given field configuration $\varphi$ to a number. The Poisson bracket is defined on the space of such functionals. An important example of such a functional is obtained by integrating $\varphi$ against some bulk source $\tilde{\lambda}$ (we write the bulk source with a tilde to distinguish it from the boundary source $\lambda$):
\beq 
\mathcal{F} = \int dX \, \widetilde{\lambda}(X)\varphi(X)~.
\eeq
Given a functional $\mathcal{F}$, one can define a vector field $\mathcal{X}_{\mathcal{F}}$ that satisfies the condition 
\beq \label{eq:Hamiltonianvf}
i_{\mathcal{X}_{\mathcal{F}}}\Omega = \delta \mathcal{F}~,
\eeq
which is known as a Hamiltonian vector field. The Poisson bracket associated to the symplectic form is now defined by the formula
\beq \label{eq:Poissonbracket}
\{\mathcal{F},\mathcal{G}\} \equiv \Omega(\mathcal{X}_{\mathcal{F}}, \mathcal{X}_{\mathcal{G}})~,
\eeq 
where $\mathcal{X}_{\mathcal{F}}$ and $\mathcal{X}_{\mathcal{G}}$ are the Hamiltonian vector fields associated to the functionals $\mathcal{F}$ and $\mathcal{G}$ with respect to $\Omega$. We would like to argue that the Poisson bracket derived from the bulk symplectic form \eqref{eq:covsymform}, for a suitable choice of functionals $\mathcal{F}$ and $\mathcal{G}$, computes the Berry curvature $F_{\Psi}$ on the boundary.

\subsection{Berry curvature = bulk symplectic form} \label{sec:Berry = bulk}

In this section, we will go through the details of this argument, establishing a direct link between the modular Berry phase and the bulk covariant phase space symplectic form. The first step is to extend \eqref{eq:FOmega} into the bulk. Recall that given a point $X\in D_R$ in the right exterior region, one can express a local bulk operator $\phi(X)$ as an integral over the boundary 
\beq \label{eq:bulkrecon}
\phi(X) = \int dx \, K(x|X)\mathcal{O}(x)~,
\eeq
using the so-called bulk-to-boundary propagator $K(x|X)$. This is a version of the HKLL bulk reconstruction map \cite{Hamilton:2006az, Hamilton:2006fh}. The operator $\mathcal{O}(x)$ is obtained from $\phi(X)$ by taking a boundary limit in the sense of \eqref{eq:extrapolate}. The bulk-to-boundary propagator can be formally expressed as an integral over the bulk Fourier modes, but in practice this is quite subtle. In many examples, the large $q$ behavior of the modes $\alpha_{\omega,q}$ makes the corresponding integral divergent, and consequently the bulk-to-boundary propagator $K(x|X)$ is not a well-defined function \cite{Morrison:2014jha} (see also \cite{Bousso:2012mh, Papadodimas:2012aq}). To avoid divergences, one can smear the local bulk operator $\phi(X)$ against some test function with suitable large $q$ behavior. An alternative solution is to implement the bulk-to-boundary map at fixed quantum number $q$. For the present purposes, we simply understand all appearances of the causal propagator to be ultimately smeared against some suitable test function regularizing the final result.

Let us come back to the expression \eqref{eq:FOmega} for the modular Berry curvature. Using the GFF property \eqref{eq:comm} combined with bulk reconstruction map \eqref{eq:bulkrecon}, one can express the commutator of two fundamental bulk fields as
\beq \label{eq:commbulk}
[\phi(X),\phi(Y)]= -i \int dx \int dy \, K(x|X)K(y|Y) \propegator(x-y)\,\mathds{1}~.
\eeq
One can read off from \eqref{eq:commbulk} that the bulk causal propagator, denoted by $\widetilde{\propegator}$, can be expressed in terms of the boundary one via the integral transform
\beq 
\widetilde{\propegator}(X,Y)= \int dx \int dy \, K(x|X)K(y|Y) \propegator(x-y)~,
\eeq
and \eqref{eq:commbulk} is now simply the statement that the bulk field $\phi$ is also described by a GFF theory. If we now smear the bulk-to-boundary propagator against some suitable bulk perturbation $\delta_1 \widetilde{\lambda}$ and $\delta_2 \widetilde{\lambda}$, the corresponding boundary perturbation takes the form 
\beq 
\delta_1 \lambda(x) = \int dX\, K(x|X) \delta_1 \widetilde{\lambda}(X)~, \qquad \delta_2 \lambda(y) = \int dY\, K(y|Y) \delta_2 \widetilde{\lambda}(Y)~.
\eeq
It follows that the bulk expression for the modular Berry curvature can be written as
\beq \label{eq:bulkmodberry}
F_{\Psi}= i \int dX \int dY\,\delta_1\widetilde{\lambda}(X)\widetilde{\propegator}(X,Y)\delta_2\widetilde{\lambda}(Y)~.
\eeq

To see how this connects to the bulk symplectic form, we need to find an explicit expression for the Hamiltonian vector fields. Written out explicitly in terms of variational derivatives, the defining equation \eqref{eq:Hamiltonianvf} of the Hamiltonian vector field is given by 
\beq \label{eq:OmegaXF}
\Omega(\mathcal{X}_{\mathcal{F}}, \mathcal{Y}) = \int dX \frac{\delta \mathcal{F}}{\delta \varphi(X)}\mathcal{Y}(X)~,
\eeq
where $\mathcal{Y}$ is an arbitrary vector field. To find $\mathcal{X}_{\mathcal{F}}$ in the above equation, we introduce the \emph{Jacobi operator} that implements the linearized equation of motion. As before, we will assume to be working with a massive free scalar field on a background geometry given by a metric $g_{\mu\nu}$, which satisfies the usual Klein-Gordon field equation. In that case, the Jacobi operator takes the form $\mathfrak{J}(X,Y)=\delta(X-Y)\mathfrak{J}$, where the differential operator $\mathfrak{J}$ can be written as (see, e.g., \cite{Gieres:2021ekc}) 
\beq \label{eq:Jfree}
\mathfrak{J}_{IJ} = \begin{pmatrix} -m^2 & -\nabla_{\nu} \\ 
\nabla_{\mu} & -g_{\mu\nu}
\end{pmatrix}~.
\eeq
As before, the label $I$ (and $J$) runs over the field and canonical momentum variables via $\varphi^I=(\varphi,\pi^{\mu})$. Imposing $\mathfrak{J}(X,Y)_{IJ}\delta\varphi^J = 0$ leads to the following first order form of the field equations
\beq \label{eq:on-shellrelation}
\delta \pi_{\mu} = \nabla_{\mu}\delta \varphi~,\qquad \nabla_{\mu}\delta \pi^{\mu}+m^2\delta \varphi = 0~.
\eeq 
We now introduce the retarded and advanced Green functions, denoted by  $\widetilde{\propegator}^{-}, \widetilde{\propegator}^{+}$ respectively, as the inverse of the Jacobi operator in the sense that 
\beq \label{eq:defGreenf}
\int d Y \, \mathfrak{J}(X,Y)_{IJ} \widetilde{\propegator}^{\pm}(Y,Z)^{JK} = -\delta_I^{K}\delta(X-Z)~.
\eeq
The causal Green function is now defined as the difference between retarded and advanced Green functions:
\beq 
\widetilde{\propegator}\equiv \widetilde{\propegator}^{-}-\widetilde{\propegator}^{+}~.
\eeq
Here, we have extended the notation of the bulk causal propagator, cf. \eqref{eq:defGreenf}, to include the conjugate momentum operators $\Pi^{\mu}$ as well. Writing $\phi^{I}=(\phi,\Pi^{\mu})$ as we did for the classical counterpart, the causal propagator appears in the commutator of the fundamental bulk operator and conjugate momentum operators as $[\phi^{I}(X),\phi^J(Y)]= -i \widetilde{\propegator}(X,Y)^{IJ}\mathds{1}$. 

Given a functional $\mathcal{F}$ on phase space, let us now introduce the vector fields $\mathcal{X}_{\mathcal{F}}^{\pm}$ with components defined by 
\beq
(\mathcal{X}_{\mathcal{F}}^{\pm})^I \equiv \int d Y \, \widetilde{\propegator}^{\pm}(X,Y)^{IJ} \frac{\delta \mathcal{F}}{\delta \varphi^J(X)}~.
\eeq
By definition of the retarded and advanced Green functions \eqref{eq:defGreenf}, they satisfy the equation
\beq \label{eq:inhom}
\int dY\, \mathfrak{J}(X,Y)_{IJ} (\mathcal{X}_{\mathcal{F}}^{\pm
})^J(Y) = - \frac{\delta \mathcal{F}}{\delta \varphi^I(X)}~.
\eeq
We plug this vector field into the current $J^{\mu}$ in \eqref{eq:current} that is obtained from the covariant phase space method:
\beq 
J^{\mu}(\mathcal{X}_{\mathcal{F}}^{\pm}, \delta \varphi) = (\mathcal{X}_{\mathcal{F}}^{\pm})^{\varphi}\delta \pi^{\mu}-\delta \varphi (\mathcal{X}_{\mathcal{F}}^{\pm})^{\mu}~,
\eeq
where we have written the components of the vector field as $\mathcal{X}_{\mathcal{F}}^{\pm}=((\mathcal{X}_{\mathcal{F}}^{\pm})^{\varphi},(\mathcal{X}_{\mathcal{F}}^{\pm})^{\mu})$.
We now compute the divergence $\nabla_{\mu}J^{\mu}$ using the on-shell relations \eqref{eq:on-shellrelation} as
\begin{align} \label{eq:nablaJ}
\nabla_{\mu}J^{\mu}(\mathcal{X}_{\mathcal{F}}^{\pm}, \delta \varphi) = -\left(\nabla_{\mu}(\mathcal{X}_{\mathcal{F}}^{\pm})^{\mu}+ m^2 (\mathcal{X}_{\mathcal{F}}^{\pm})^{\varphi}\right)\delta \varphi + \left((\nabla_{\mu}\mathcal{X}_{\mathcal{F}}^{\pm})^{\varphi} -g_{\mu\nu}(\mathcal{X}_{\mathcal{F}}^{\pm})^{\nu} \right)\delta \pi^{\mu} ~.
\end{align}
Looking at the form of the Jacobi operator \eqref{eq:Jfree}, one can write \eqref{eq:nablaJ} in the form
\beq \label{eq:identity}
\nabla_{\mu}J^{\mu}(\mathcal{X}_{\mathcal{F}}^{\pm}, \delta \varphi) =(\mathfrak{J} \cdot \mathcal{X}_{\mathcal{F}}^{\pm})_{J}  \delta \varphi^J = -\frac{\delta \mathcal{F}}{\delta \varphi^J}\,\delta \varphi^J ~.
\eeq
Next, we integrate the above identity to get an expression for the symplectic form. Let us introduce two spatial surfaces $\Sigma^{+}$ and $\Sigma^{-}$, to the future and past of the initial slice $\Sigma$ respectively, such that the vector fields vanish
\beq \label{eq:boundarycondition}
\mathcal{X}_{\mathcal{F}}^{\pm}|_{\Sigma^\pm}=0~,
\eeq
as follows from basic properties of the retarded and advanced Green functions. Moreover, we assume that the perturbation $\delta \mathcal{F}$ only has support in the region that is bounded by the two surfaces $\Sigma_{+}$ and $\Sigma_{-}$. The bulk region between the spatial surfaces $\Sigma$ and $\Sigma_{+}$ is denoted by $S_{+}$, while we denote the region between $\Sigma_{-}$ and $\Sigma$ by $S_{-}$. Let us now integrate equation \eqref{eq:identity} for $\mathcal{X}_{\mathcal{F}}^{+}$ over the upper region $S_+$. Because the left-hand side of \eqref{eq:identity} is a total derivative, using Stokes' theorem we can rewrite this expression as a boundary integral:
\begin{align}
\int_{S_{+}} dX\,\frac{\delta \mathcal{F}}{\delta \varphi^I} \, \delta \varphi^I = - \int_{\Sigma_{+}} d\Sigma_{\mu}\, J^{\mu}(\mathcal{X}_{\mathcal{F}}^{+},\delta\varphi)+\int_{\Sigma} d\Sigma_{\mu}\, J^{\mu}(\mathcal{X}_{\mathcal{F}}^{+},\delta\varphi)~.
\end{align}
Similarly, taking instead the vector field $\mathcal{X}_{\mathcal{F}}^{-}$, and integrating over the lower region $S_-$ one finds that 
\begin{align}
\int_{S_{-}} dX\,\frac{\delta \mathcal{F}}{\delta \varphi^I}\,\delta \varphi^I =- \int_{\Sigma} d\Sigma_{\mu} \, J^{\mu}(\mathcal{X}_{\mathcal{F}}^{-},\delta\varphi)+\int_{\Sigma_{-}} d\Sigma_{\mu}\, J^{\mu}(\mathcal{X}_{\mathcal{F}}^{-},\delta\varphi)~.
\end{align}
The conditions \eqref{eq:boundarycondition} imply that the integrals over $\Sigma_{+}$ and $\Sigma_{-}$ in the above expression vanish. Therefore, adding both equations and introducing the difference of retarded and advanced degrees of freedom 
\beq \label{eq:deltaF}
\mathcal{X}_{\mathcal{F}}= \mathcal{X}_{\mathcal{F}}^- - \mathcal{X}_{\mathcal{F}}^+ =  \int dY\, \widetilde{\propegator}(X,Y) \frac{\delta \mathcal{F}}{\delta \varphi(X)}~,
\eeq 
one finds that 
\beq 
\int_{S_{-}\cup S_{+}} dX\,\frac{\delta 
\mathcal{F}}{\delta \varphi^I}\,\delta \varphi^I = \int_{\Sigma} d\Sigma_{\mu}\,J^{\mu}(\mathcal{X}_{\mathcal{F}},\delta\varphi)~.
\eeq
The right-hand side of this equation is the bulk symplectic form. By assumption the integrand on the left-hand vanishes outside of the region $S_{-}\cup S_{+}$, so we can replace the integration domain by the full right wedge $D_{R}$ without changing the outcome. We conclude that  
\beq 
\Omega(\mathcal{X}_{\mathcal{F}}, \delta \varphi)= \int dX \, \frac{\delta \mathcal{F}}{\delta \varphi^I}\, \delta \varphi^I~,
\eeq
which is precisely the identity \eqref{eq:OmegaXF} that needs to be satisfied for $\mathcal{X}_{\mathcal{F}}$ to be the Hamiltonian vector field associated to the functional $\mathcal{F}$. 

Given the explicit form \eqref{eq:deltaF} of the Hamiltonian vector field, we can take $\mathcal{Y}=\mathcal{X}_{\mathcal{G}}$ in \eqref{eq:OmegaXF}. Plugging the expression for $\mathcal{X}_{\mathcal{G}}$ into the symplectic form the final result is given by 
\beq \label{eq:Peierlsbracket}
\{\mathcal{F},\mathcal{G}\}= \int dX\, \frac{\delta \mathcal{F}}{\delta \varphi^I(X)} \mathcal{X}_{\mathcal{G}}^I(X) = \int dX \int dY \, \frac{\delta \mathcal{F}}{\delta \varphi^I(X)} \widetilde{\propegator}(X,Y)^{IJ}\frac{\delta \mathcal{G}}{\delta \varphi^J(Y)}~.
\eeq
The right-hand side of \eqref{eq:Peierlsbracket} is actually known as the \emph{Peierls bracket}, supported on a \emph{spacetime} region, as opposed to the usual representation of the Poisson bracket on a spatial slice, which is related to the equal-time commutator bracket. Implicitly, we have used the dynamics of the theory to extend the result to the whole right spacetime region. To connect to the expression for the modular Berry curvature $F_{\Psi}$ in \eqref{eq:bulkmodberry}, we take the functionals in the above equation to be of the form 
\beq 
\mathcal{F}=\int dX\, \delta_1 \widetilde{\lambda}(X) \varphi(X)~, \qquad \mathcal{G}=\int dY\, \delta_2 \widetilde{\lambda}(Y) \varphi(Y)~,
\eeq
where $\delta_1 \widetilde{\lambda}$ and $\delta_2 \widetilde{\lambda}$ are two bulk perturbations. The Peierls bracket in the above expression then reduces to 
\beq 
\{\mathcal{F},\mathcal{G}\} = \int dX\int dY\,\delta_1\widetilde{\lambda}(X)\widetilde{\propegator}(X,Y) \delta_2\widetilde{\lambda}(Y) = -i F_{\Omega}~.
\eeq
Hence, we have established our anticipated relation 
\beq 
F_{\Psi}= i\, \Omega(\mathcal{X}_{\mathcal{F}},\mathcal{X}_{\mathcal{G}})~,
\eeq
which is the main result of the present section. It provides another instance where the modular Berry curvature, in this case for the large $N$ von Neumann algebra associated to the TFD state, can be related to a symplectic form in the dual bulk geometry. 

Let us end with two remarks. Firstly, we point out that the parallel transport condition $P_0(X)=0$ in the Berry curvature computation should fix the boundary condition at the bifurcation surface to be Dirichlet. Indeed, the operators localized on the bifurcation surface are fixed by the modular flow, so from the boundary perspective they should correspond to modular zero modes. We refer to Section 4 of \cite{Faulkner:2017vdd} for a formula expressing the zero mode $P_0(\mathcal{O})$ in terms of an integral of the dual bulk field over the bifurcation surface. As was already explained in \cite{Czech:2023zmq}, it now follows that projecting out the zero mode in the boundary perturbation \eqref{eq:paralleltransport} makes the terms that are localized at the bifurcation surface vanish. This amounts to choosing Dirichlet boundary conditions for the symplectic form $\Omega$.

Secondly, it is important to remark that the result derived above is expected to be true in more general settings. While we have performed our computation in the context of the TFD state dual to an eternal black hole and considered the full right boundary, we expect the derivation to go through when we consider a more general class of states and subregions. A crucial assumption in the boundary computation of the Berry curvature is that the correlation functions are well-described by a GFF. This allows us to express the result in simple terms, and make a connection with the Peierls bracket. For general subregions in the boundary CFT, the corresponding bulk symplectic form should be associated to the corresponding entanglement wedge (see also the results of \cite{Czech:2023zmq}).   

\subsection{Comments on the K\"ahler structure of phase space}

We have shown that the Berry curvature provides a natural symplectic form on the phase space of the theory. We will now argue that there is a compatible metric and almost complex structure. Indeed, if we write out the commutator of $\mathcal{O}(x)$ with $\mathcal{O}(y)$ using the mode expansion \eqref{eq:bdymodeexpansion} and canonical commutation relations \eqref{eq:canonicalcom}, one finds that the boundary causal propagator can be written as 
\beq \label{eq:Deleta}
\propegator(x-y)=\int dq\int_0^{
\infty} \frac{d\omega}{2\pi}(\gamma_{\omega,q}(x)\gamma_{\omega,q}^*(y)- \gamma_{\omega,q}(y)\gamma_{\omega,q}^*(x))~.
\eeq
Plugging this into \eqref{eq:FOmega}, and introducing the fixed frequency modes
\beq 
\delta\widehat{\lambda}_{\omega,q} = \int dx \, \gamma_{\omega,q}(x) \delta \lambda(x)~, \qquad \delta\widehat{\lambda}_{\omega,q}^{*} = \int dx \, \gamma_{\omega,q}^*(x) \delta \lambda(x)~,
\eeq
one obtains the following expression for the Berry curvature:
\beq 
F_{\Psi} =i \int dq \int_0^{
\infty} \frac{d\omega}{2\pi}\left( \delta_1\widehat{\lambda}_{\omega,q}^*\delta_2\widehat{\lambda}_{\omega,q} - \delta_1\widehat{\lambda}_{\omega,q}\delta_2\widehat{\lambda}_{\omega,q}^*\right)~.
\eeq
This is the representation of the symplectic form in momentum space. It turns out that there is a natural K\"ahler structure associated to the symplectic form. To make this manifest, we point out the Hermitian inner product in \eqref{eq:hermitianmetric} can be decomposed as 
\beq 
\langle\delta_1\lambda|\delta_2\lambda \rangle = g(\delta_1\lambda,\delta_2\lambda)+i\omega(\delta_1\lambda,\delta_2\lambda)~,
\eeq
where we have introduced the metric as the real part of the inner product 
\beq 
g(\delta_1\lambda,\delta_2\lambda)\equiv \mathfrak{Re}\,\langle\delta_1\lambda|\delta_2 \lambda \rangle = \frac{1}{2}\int dq\int_0^{
\infty} \frac{d\omega}{2\pi}\left( \delta_1\widehat{\lambda}_{\omega,q}^*\delta_2\widehat{\lambda}_{\omega,q} + \delta_1\widehat{\lambda}_{\omega,q}\delta_2\widehat{\lambda}_{\omega,q}^*\right) ~,
\eeq
and the symplectic form as the imaginary part
\beq \label{eq:sympfrequency}
\omega(\delta_1\lambda,\delta_2\lambda) = \mathfrak{Im}\,\langle\delta_1\lambda|\delta_2\lambda \rangle = \frac{1}{2i}\int dq\int_0^{
\infty} \frac{d\omega}{2\pi}\left( \delta_1\widehat{\lambda}_{\omega,q}^*\delta_2\widehat{\lambda}_{\omega,q} - \delta_1\widehat{\lambda}_{\omega,q}\delta_2\widehat{\lambda}_{\omega,q}^*\right)~.
\eeq
Note that the symplectic form introduced in this way is proportional to the modular Berry curvature in the sense that $-2F_{\Psi}=\omega$. Clearly, if we now introduce the almost complex structure $J$ that implements the multiplication by $i$ on the source functions,
\beq 
J(\delta\lambda)\equiv i\delta\lambda~,
\eeq
both the metric and the symplectic form are compatible in the sense that 
\beq
g(\delta_1\lambda,\delta_2\lambda)=\omega(\delta_1\lambda,J(\delta_2\lambda))~.
\eeq
This proves that the phase space with the inner product defined in \eqref{eq:hermitianmetric} is a K\"ahler manifold. The corresponding symplectic form is given by the modular Berry curvature. Such a K\"ahler structure on the space of pure state deformations from the Berry phase was also made apparent in other works, see e.g., \cite{Belin:2018fxe, Belin:2018bpg}.

%% file: OTOC.tex
\section{Modular chaos and shape-changing deformations}

\label{sec:modchaos}

Up to now we have focused on the state-perturbations coming from unitaries that are in the algebra of operators. Interestingly, the framework of the modular Berry phase supports a broader class of perturbations including so-called \emph{shape-deformations}. These correspond to transformations that leave the global state of the theory fixed, but change the shape of a subregion on the boundary. For boundary CFTs in the vacuum state, where the shape-changes are generated by symmetry transformations, this leads to the structure of kinematic space. From the algebraic perspective, one needs to consider \emph{global} transformations that act nontrivially on the full algebra of observables, while leaving the cyclic and separating state fixed. 

An important example that illustrates the general idea involves the so-called \emph{half-sided modular inclusion}, which has received some attention lately. The concept of a half-sided modular inclusion and its associated group structure was first introduced by Wiesbrock \cite{Wiesbrock:1992mg,Araki-Zsido}, based on earlier work by Borchers \cite{Borchers:1991xk}. Wiesbrock showed that the half-sided modular inclusion property can be characterized by the existence of a one-parameter group of unitaries with positive Hermitian generator, which is governed by exponential growth under modular time evolution (see \cite{Borchers:2000pv} for a review). Physically, the exponential growth of modular perturbations is interpreted as a manifestation of the system's underlying quantum chaos, dubbed \emph{modular chaos}. Specifically, in \cite{DeBoer:2019kdj} it was argued that the growth of such perturbations is bounded, and that the associated modular chaos bound is saturated in the case of a half-sided modular inclusion. One can therefore understand the algebraic structure as representing a version of maximal modular chaos, an idea that has been further elaborated on in \cite{Gesteau:2023rrx, Ouseph:2023juq} by exhibiting an underlying ergodic hierarchy. 

In the present section, we would like to add an additional ingredient to the discussion by arguing that the above connections can be understood in terms of a suitable computation of the modular Berry curvature. In particular, this provides a direct link between the modular structure of the theory and its underlying quantum chaotic properties. Making this precise  requires an extension of our framework to include global transformations. We will first review the definition of a half-sided modular inclusion, and the associated \emph{modular scrambling modes}. We will then argue that the vanishing of the modular Berry curvature for modular scrambling modes implies the existence of an emergent Poincar\'e algebra, which is presented holographically in the local near-horizon geometry of the bulk dual. To make the connection with quantum chaos more manifest, we will define a modular version of the out-of-time-ordered correlator (OTOC), and show that it can be computed in terms of the Berry phase for a suitable choice of global perturbations.

\subsection{Half-sided modular inclusions and modular scrambling modes}

Let us consider an inclusion of von Neumann algebras $\mathcal{B}\subset \mathcal{A}$ with $|\psi\rangle$ a common cyclic and separating vector for both algebras. We will denote the corresponding modular operators associated to the algebras $\mathcal{B}$ and $\mathcal{A}$ by $\Delta_{\mathcal{B}}$ and $\Delta_{\mathcal{A}}$, respectively. The inclusion $\mathcal{B}\subset \mathcal{A}$ is called a future/past \emph{half-sided modular inclusion} with respect to the state $|\psi\rangle$ if the following condition is satisfied:
\beq \label{eq:HSMI}
\Delta^{-is}_{\mathcal{A}}\mathcal{B}\Delta^{is}_{\mathcal{A}} \subseteq \mathcal{B}~, \qquad \pm s \geq 0~,
\eeq
which simply says that the smaller algebra $\mathcal{B}$ is mapped into itself by the modular flow associated to the larger algebra $\mathcal{A}$ for either positive (future) or negative (past) modular times, but not both. Let us first focus on future half-sided modular inclusions, while keeping in mind that every statement has a counterpart for past half-sided modular inclusions as well.

An important consequence of the modular inclusion property \eqref{eq:HSMI}, as was proved in \cite{Araki-Zsido, Borchers:1995zg}, is the existence of a positive self-adjoint operator $G_+$, which can be expressed in terms of the modular operators as 
\beq \label{eq:G+}
G_{+} \equiv \frac{1}{2\pi} \left( \log \Delta_{\mathcal{B}} - \log \Delta_{\mathcal{A}}\right)~.
\eeq
The above generator has a number of interesting properties. Notably, the one-parameter family of unitaries that is generated by $G_+$,
\beq \label{eq:u+}
\mathfrak{u}_+(t) \equiv e^{itG_{+}}~, \qquad t \in \mathbb{R}~,
\eeq
can be thought of as a continuous `translation' on the algebra that can be used to map $\mathcal{A}$ to its subalgebra $\mathcal{B}$. In particular, one can show that it satisfies $\mathfrak{u}_+(t)\mathcal{A}\mathfrak{u}_+(t)^{\dagger} \subseteq \mathcal{A}$ for $t \geq 0$ with the property that $\mathcal{B} = \mathfrak{u}_+(1)\mathcal{A}\mathfrak{u}_+(1)^{\dagger}$. In addition, the operator $\mathfrak{u}_+(t)$ has the following behavior under modular flow of both algebras:
\beq \label{eq:algebra}
\Delta^{-is}_{\mathcal{A}}\mathfrak{u}_+(t)\Delta^{is}_{\mathcal{A}} = \Delta^{-is}_{\mathcal{B}}\mathfrak{u}_+(t)\Delta^{is}_{\mathcal{B}} = \mathfrak{u}_+(e^{2\pi s} t)~.
\eeq 
Note that $G_+|\psi\rangle =0$, as can be seen from \eqref{eq:G+}. Consequently, the unitaries $\mathfrak{u}_+(t)$ leave the global state invariant, while at the same time having a non-trivial action on the algebra. This is somewhat different from the situation that was considered in previous sections, where we studied perturbations that explicitly change the state. When the algebras in question are associated to subregions in spacetime, the above transformations can be associated with changes in the shape of the subregion leaving the global state invariant. For this reason, we will refer to them as \emph{shape-changing} transformations to distinguish them from the state-changing ones. The separating property of the state $|\psi\rangle$ implies that the operator $G_+$ cannot be localized within the algebra $\mathcal{A}$.

\begin{figure}
    \centering
    \includegraphics[width=0.5\linewidth]{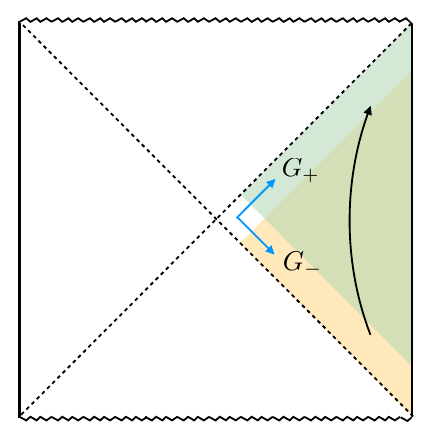}
    \caption{A half-sided modular inclusion in the eternal black hole geometry. We have indicated two subalgebras $\mathcal{B}_+$ (green) and $\mathcal{B}_{-}$ (yellow) which constitute respectively a future and past half-sided modular inclusion with respect to the modular flow (black arrow) of the right algebra $\widetilde{\mathcal{A}}_R$. The modular scrambling modes $G_{+}$ and $G_{-}$ act geometrically in the near-horizon region as (approximate) null translations (indicated by blue arrows).}
    \label{fig:HSMI}
\end{figure}

A geometrical example of the half-sided modular inclusion can be given in Rindler space, if we choose the larger algebra $\mathcal{A}$ to be associated to the right wedge, and the subalgebra $\mathcal{B}$ to be a slightly smaller nested wedge that shares a null boundary with $\mathcal{A}$. Because the modular flow of $\mathcal{A}$ acts as a boost in the right wedge, it automatically maps the smaller algebra into itself for positive modular times. Therefore, $\mathcal{B}\subset\mathcal{A}$ constitutes a half-sided modular inclusion. We can now understand the generator $G_+$ in \eqref{eq:G+} in terms of the composition of two boosts, corresponding to the larger and smaller algebra respectively. The result is a null translation that shifts the algebra $\mathcal{A}$ into $\mathcal{B}$. This example can be generalized to more general spacetimes with a bifurcate horizon, while the operator $G_+$ with the required properties has a nice geometrical representation in terms of a null translation only when restricted to the horizon, see Figure \ref{fig:HSMI}.

The commutation relation in \eqref{eq:algebra} can be recast in terms of the modular Hamiltonians associated to both algebras.
Note that the operator $G_+$ can be written as the difference of modular Hamiltonians:
\beq \label{eq:definitionG+}
K_{\mathcal{B}}= - \log \Delta_{\mathcal{B}}~, \qquad  K_{\mathcal{A}}= - \log \Delta_{\mathcal{A}}, \qquad G_+ = \frac{1}{2\pi}\left( K_{\mathcal{A}} - K_{\mathcal{B}}\right)~.
\eeq
The algebra relations \eqref{eq:algebra} in terms of the above generators now take the simple form:
\beq \label{eq:commutationG+}
[K_{\mathcal{A}},G_+]= [K_{\mathcal{B}},G_+]= 2\pi i G_{+}~.
\eeq
There is a similar statement for past half-sided modular inclusions. The resulting translation operator is denoted by $G_-$, and the commutation relations with the modular Hamiltonian are identical to the ones in \eqref{eq:commutationG+}, except for an extra minus sign on the right-hand side of the equation:
\beq \label{eq:commutationG-}
[K_{\mathcal{A}}, G_-]= [K_{\mathcal{B}},G_-]=-2\pi i G_-~.
\eeq 
The relation with chaos comes from the \emph{modular chaos bound} that was derived in \cite{DeBoer:2019kdj}, which provides an upper bound on the growth of perturbations of the modular Hamiltonian under modular flow. As the operators $G_{+}$ and $G_{-}$ saturate the modular chaos bound, they are in some appropriate sense `maximally chaotic', and are referred to as \emph{modular scrambling modes}.

\subsection{Vanishing of the Berry phase implies an emergent Poincar\'e algebra}

We have so far considered Berry phases associated to state perturbations of $|\psi\rangle$ that arise from applying a one-parameter family of unitaries $u(\lambda)$ in the algebra $\mathcal{A}$. However, to accommodate for shape-changing transformations, one has to extend the formalism somewhat and allow for unitaries that are in the global algebra of observables. We will refer to such unitaries as \emph{global perturbations}, as they act on both $\mathcal{A}$ and $\mathcal{A}'$. By definition, the von Neumann algebra $\mathcal{A} \vee \mathcal{A}' \equiv (\mathcal{A}\cup \mathcal{A}')''$ generated by the algebra $\mathcal{A}$ and its commutant $\mathcal{A}'$, is given by the algebra $B(\mathcal{H})$ of all bounded operators on the Hilbert space $\mathcal{H}$. Examples of such global unitaries include the operator $\mathfrak{u}_{+}(t)$ in \eqref{eq:u+} and the operator $\Delta^{is}$\footnote{Recall that in general the modular operator $\Delta$ is not bounded, so it can only be defined on a dense subset of the full Hilbert space. As a consequence, $\Delta^{is}$ is a priori defined on the same domain. Its action can however be extended to the full Hilbert space, and therefore $\Delta^{is} \in B(\mathcal{H})$.}, which both have a non-trivial action on the commutant algebra $\mathcal{A}'$. 

To explain the idea of the shape-changing parallel transport, we will go back to the example of half-sided modular inclusions. In order for a non-trivial geometric phase to come out, we will assume the existence of both translation generators $G_+$ and $G_-$ with the commutation relations \eqref{eq:commutationG+} and \eqref{eq:commutationG-}. This corresponds to the setting of a simultaneous $-$ and a $+$ half-sided modular inclusion. We can now use the unitaries 
\beq \label{eq:U+U-}
\mathfrak{u}_+(t) = e^{it G_{+}}~, \qquad \mathfrak{u}_-(t)=e^{itG_-}~, \qquad t\in \mathbb{R}~,
\eeq
to introduce a notion of parallel transport. Note however that the transformations in \eqref{eq:U+U-} do not change the state $|\psi\rangle$. Indeed, as was stressed before the generators $G_{\pm}$ satisfy $G_{\pm}|\psi\rangle =0$, from which it follows that 
\beq 
\mathfrak{u}_{\pm}(t)|\psi\rangle = |\psi\rangle~.
\eeq
Because the global state is left invariant, it naively seems that there is no sensible notion of parallel transport associated to this class of perturbations. However, the global unitaries do change the algebra operators. For example, using the generator $G_+$, one can define the one-parameter family of algebras given by 
\beq 
\mathcal{A}_{t,+}\equiv \mathfrak{u}_+(t)\mathcal{A}\mathfrak{u}_+(t)^{\dagger}~.
\eeq 
In particular, the modular operator for the deformed algebra gets mapped to
\beq 
\Delta_{\mathcal{A}}\to \mathfrak{u}_+(t)\Delta_{\mathcal{A}}\mathfrak{u}_+(t)^{\dagger}~.
\eeq
This defines a path in the space of modular Hamiltonians. The above observation allows us to still define a non-trivial \emph{modular} Berry transport for shape deformations, even though the global state is kept fixed. We will refer to this as the ``two algebras, one state'' scenario. The idea is to make a small loop by a suitable concatenation of paths. To achieve this, we allow perturbations of the modular operator in a second direction by using the generator $G_-$. We now construct a closed loop by applying a translation with $G_+$ over a small distance $\varepsilon$, then acting with $G_-$ over the same distance, and going back in the opposite order. By Hermiticity of the operator $G_+$, the Hermitian conjugates of \eqref{eq:U+U-} lead to the same paths but with the opposite direction $t\to -t$. This leads to the following expression:
\beq \label{eq:loop}
\mathfrak{u}_+(\varepsilon)\mathfrak{u}_-(\varepsilon)\mathfrak{u}_+(\varepsilon)^{\dagger}\mathfrak{u}_-(\varepsilon)^{\dagger}~.
\eeq
Taking $\varepsilon \to 0$ in the above expression, will probe the modular Berry curvature on the space of deformations. To be precise, the second order term in $\varepsilon$-expansion of the loop \eqref{eq:loop} corresponds to the curvature. As explained in Section \ref{sec:algebraBerry}, one needs to project the result into the space of zero modes. In the present case, it is straightforward to check how the operator \eqref{eq:loop} behaves under modular evolution. Using the commutation relations in \eqref{eq:algebra}, one finds that
\begin{align} \label{eq:computation}
\Delta^{-is}&\left( \mathfrak{u}_+(\varepsilon)\mathfrak{u}_-(\varepsilon)  \mathfrak{u}_+(\varepsilon)^{\dagger}\mathfrak{u}_-(\varepsilon)^{\dagger}\right) \Delta^{is} 
= \mathfrak{u}_+(e^{2\pi s}\varepsilon)\mathfrak{u}_-(e^{-2\pi s}\varepsilon)\mathfrak{u}_+(e^{2\pi s}\varepsilon)^{\dagger}\mathfrak{u}_-(e^{-2\pi s}\varepsilon)^{\dagger} \nonumber \\
&=e^{ie^{-2\pi s}\varepsilon G_-- \varepsilon^2[G_+,G_-]+\cdots}\,\mathfrak{u}_-(e^{-2\pi s}\varepsilon)^{\dagger} \nonumber =e^{ie^{-2\pi s}\varepsilon G_-- \varepsilon^2[G_+,G_-]+\cdots}\,e^{-ie^{-2\pi s}\varepsilon G_{-}}\nonumber \\
&=e^{-\varepsilon^2 [G_+,G_-]+\cdots} = 1+\varepsilon^2 [G_+,G_-]+\cdots~,
\end{align}
where we have collected in $\cdots$ all terms of order $\mathcal{O}(\varepsilon^3)$. We have used the BCH formula at the second equality to rewrite
\begin{align} 
\mathfrak{u}_+(e^{2\pi s}\varepsilon)\mathfrak{u}_-(e^{-2\pi s}\varepsilon)\mathfrak{u}_+(e^{2\pi s}\varepsilon)^{\dagger}&=e^{ie^{2\pi s} \varepsilon G_{+}}e^{ie^{-2\pi s}\varepsilon G_-} e^{-ie^{2\pi s} \varepsilon G_{+}} \nonumber \\
&= e^{ie^{-2\pi s}\varepsilon G_-- \varepsilon^2[G_+,G_-]+\cdots}~,
\end{align}
and at the fourth equality to combine two exponentials into one. Equation \eqref{eq:computation} shows that the curvature term does not depend on the variable $s$. We conclude that the Berry curvature for modular scrambling modes is given by 
\beq \label{eq:F=[G+,G-]}
F = [G_+,G_-]~.
\eeq
Note that the parallel transport equation $P_0(G_+)=P_0(G_-)=0$ is automatically satisfied in the present case due to \eqref{eq:commutationG+} and \eqref{eq:commutationG-} (see also Appendix \ref{app:A} for an independent check). Furthermore, one can check directly that the commutation relations imply that $F$ commutes with the modular Hamiltonian. Indeed, using the Jacobi identity one finds that 
\beq 
[K_{\mathcal{A}},F]=[G_+,[K_{\mathcal{A}},G_-]]-[G_-,[K_{\mathcal{A}},G_+]]=-2\pi i [G_+,G_-]-2\pi i [G_-,G_+]=0~,
\eeq
which shows that the commutator $[G_+,G_-]$ is a global modular zero mode. In the present example, one therefore does not have to apply an additional zero mode projection $P_0$ to the commutator to obtain the curvature. 

What physical information is contained in the Berry curvature? As can be seen from \eqref{eq:commutationG+} and \eqref{eq:commutationG-}, the generators $G_+$ and $G_-$ have commutation relations with the modular Hamiltonian that are the same as those appearing in the Poincar\'e algebra for the null translations and boosts, the only difference being that the generators $G_+$ and $G_-$ do not necessarily commute. The non-commutativity of the modular scrambling modes is precisely captured by the modular Berry curvature, and the existence of a Poincar\'e algebra for these generators is therefore quantified in terms of this curvature. In the context of holography, the boundary generators $G_+$ and $G_-$ can sometimes be realized geometrically in the bulk. For a certain class of AdS spacetimes with a bifurcation surface, the abstract Poincar\'e algebra on the boundary is then identified with the Poincar\'e algebra of spacetime symmetries associated to the bulk bifurcation surface \cite{DeBoer:2019kdj, Lashkari:2024lkt}. Because the existence of half-sided modular inclusions, and therefore the existence of the generators $G_{\pm}$, relies on the type III$_1$ nature of the boundary algebra, the Poincar\'e algebra is referred to as being emergent: It only arises in the large $N$ limit of the boundary theory. We summarize:
\begin{quotation}
\noindent \textit{A vanishing Berry curvature for modular scrambling modes implies an emergent Poincar\'e algebra in the bulk.}
\end{quotation}
Although a version of this statement already appeared in \cite{DeBoer:2019kdj}, the present work provides a precise algebraic definition of the modular Berry curvature in the case of modular scrambling modes. 

Let us end this subsection with two comments. Firstly, we would like to emphasize that the modular Berry curvature \eqref{eq:F=[G+,G-]} is an operator. Usually, an interesting matrix element of this operator is the expectation value in the original state $|\psi\rangle$. For instance, for the state perturbations considered in Section \ref{sec:symform} the corresponding matrix element defines a bulk symplectic form. However, in the present example this is not the case. In fact, the expectation value of the modular Berry curvature in the state $|\psi\rangle$ vanishes:
\beq 
F_{\psi}= \langle \psi|F |\psi \rangle = \langle \psi|[G_{+},G_{-}]|\psi \rangle = 0~,
\eeq
as the generators $G_{\pm}$ both annihilate the state $|\psi\rangle$. Therefore, the modular scrambling modes constitute an example where the operator $F$ contains information that is not represented in its expectation value $F_{\psi}$. 

A second comment involves the modular zero mode projection. In the above computation, we have skipped over an important subtlety. In Section \ref{sec:algebraBerry}, we have defined the modular zero modes in terms of the subalgebra of fixed points of the modular flow $\mathcal{A}^{\psi}\subset \mathcal{A}$. We then showed the existence of a conditional expectation $\mathcal{E}_0:\mathcal{A}\to \mathcal{A}^{\psi}$ that we used to project out the zero modes. A priori, the conditional expectation can only be applied to operators in $\mathcal{A}$, so it is not guaranteed that a similar formula can be used for global transformations. A natural candidate for the space of global zero modes is the subalgebra of operators $a\in B(\mathcal{H})$ which satisfy $\sigma_s(a)=a$ for all $s\in \mathbb{R}$. These global modular zero modes are also referred to as \emph{modular conserved charges}. In particular, with respect to this definition the operator $\Delta^{is}$ constitutes a global zero mode. It is not obvious that one has a well-defined projection. One could naively apply the integral formula \eqref{eq:zeromodeprojector}, but it is not guaranteed that this leads to a sensible result (see, however, Appendix \ref{app:A} for a computation in the case of the translation operators $\mathfrak{u}_{+}$). 

There might be a way to circumvent this issue in the presence of an underlying Lie algebra structure among the generators. For example, in the usual discussion of shape-changing transformations for boundary subregions through symmetry generators (see, e.g., \cite{deBoer:2021zlm}) the zero mode projector is defined in terms of the Cartan Killing form on the Lie algebra. The additional structure of an inner product allows one to define orthogonal projectors. However, in the case of an infinite-dimensional Lie algebra, this might still not lead to an unambiguous notion of the projection operator; see, for example, Appendix C of \cite{deBoer:2021zlm} for a discussion of this subtlety in the case of the Virasoro algebra. Unfortunately, we have so far not been able to satisfyingly resolve this issue, and hope to address this in future work.

\subsection{A geometric probe for modular chaos}

We would now like to explore some consequences of the previous discussion, and argue that the modular Berry curvature can serve as an interesting probe for modular chaos. In particular, having introduced the Berry phase for global transformations, it is an interesting question whether there are situations where mixed perturbations, which involve both the algebra $\mathcal{A}$ and its commutant $\mathcal{A}'$, can give rise to non-trivial phases. 

We start with a cyclic and separating state $|\psi\rangle$ in the global Hilbert space. For concreteness, it can be useful to have in mind the setting where $|\psi\rangle$ represents the TFD state $|\Psi_{\beta}\rangle$ in the system consisting of two copies of a CFT Hilbert space, cf. \eqref{eq:TFD}. The algebra $\mathcal{A}$ is then taken to be the large $N$ von Neumann algebra $\mathcal{A}_L$ acting on the left CFT, while its commutant $\mathcal{A}'$ is the algebra $\mathcal{A}_R$. However, the quantities introduced here can be defined for general von Neumann algebras, and do not rely on this specific example. Let us first consider a unitary perturbation that is localized in the algebra $\mathcal{A}$. In the TFD example, this corresponds to applying an operator to the left CFT only. We assume that the state transforms according to
\beq 
|\psi\rangle \to v(\lambda)|\psi\rangle~, \qquad  v(\lambda)
\equiv e^{i\lambda a} ~,
\eeq
where $a\in \mathcal{A}$ and $\lambda$ is some parameter. We can make the perturbation depend on modular time by replacing the operator $a$ by the modular evolved operator, which we will denote by the short-hand notation $a_{s}\equiv \sigma_s(a)$. Similarly, one can use the commutant algebra to perturb the state $|\psi\rangle$. In the following, it is useful to identify the commutant algebra $\mathcal{A}'$ in terms of $\mathcal{A}$ using the so-called \emph{modular conjugation}. To define it, one uses the fact that the Tomita map in \eqref{eq:Spsi} can be decomposed according to
\beq
S_\psi = J_\psi \Delta_\psi^{1/2}~.
\eeq
The modular conjugation $J_\psi:\mathcal{H}\to \mathcal{H}$ is an anti-unitary map, which satisfies $J_\psi^2=1$.  It is a result of the Tomita-Takesaki theory that
\beq 
J_\psi \mathcal{A}J_\psi =\mathcal{A}'~.
\eeq 
This identity allows us to associate to each operator $a\in \mathcal{A}$, acting on the left CFT, a so-called `mirror operator', denoted by 
\beq 
\widetilde{a}\equiv J_\psi a J_\psi\in \mathcal{A}'~,
\eeq
which acts on the right CFT. The state perturbation associated to the mirror operator then takes the form
\beq 
|\psi\rangle \to \widetilde{v}(\lambda)|\psi\rangle~, \qquad  \widetilde{v}(\lambda)
\equiv e^{i\lambda \widetilde{a}} ~.
\eeq
Note that the modular Berry curvature associated to the two perturbations $v(\lambda)$ and $\widetilde{v}(\lambda)$, which lie in $\mathcal{A}$ and $\mathcal{A}'$ respectively, is zero. This simply follows from the fact that $[\widetilde{a},a]=0$.

To get a non-trivial Berry phase that involves both perturbations, we will therefore have to introduce a coupling between the algebra and its commutant. We do this by applying a global transformation $\mathfrak{u}\in B(\mathcal{H})$ of the form
\beq \label{eq:u(xi)}
\mathfrak{u}(\xi)\equiv \exp\left(i \xi \mathfrak{b}\right)\in B(\mathcal{H})~.
\eeq
Here, we have introduced the parameter $\xi$, which measures the strength of the coupling between both sides. The idea is to first consider a perturbation by the operator $a_s\in \mathcal{A}$. This corresponds to applying the $s$-dependent unitary $v_s(\lambda)\equiv e^{i\lambda a_s}$. The second perturbation involves a unitary in the commutant $\widetilde{v}_{s}(\lambda)\equiv e^{i\lambda \widetilde{a_s}}$,
which we conjugate with the unitary $\mathfrak{u}(\xi)$ to couple it to the algebra $\mathcal{A}$. Note that the mirror operators evolve in time with the opposite sign due to the identity $J_\psi \Delta_\psi J_\psi = \Delta_\psi^{-1}$. It follows that
\beq 
\widetilde{a_s} = J_{\psi}\Delta_{\psi}^{-is}a \Delta_{\psi}^{is}J_{\psi} = \Delta_{\psi}^{is} J_{\psi}a J_{\psi}  \Delta_{\psi}^{-is}= \widetilde{a}_{-s} \in \mathcal{A}'~.
\eeq 
If we expand the second perturbation up to first order in $\lambda$, we find that 
\beq 
\mathfrak{u}(\xi)\widetilde{v}_{s}(\lambda)\mathfrak{u}(\xi)^{\dagger} = 1 + i\lambda \mathfrak{u}(\xi)\widetilde{a}_{-s}\mathfrak{u}(\xi)^{\dagger}+\cdots~,
\eeq
which shows that the generator of the second perturbation is given by $\mathfrak{u}(\xi)\widetilde{a}_{-s}\mathfrak{u}(\xi)^{\dagger} \in B(\mathcal{H})$.
At this point, one would have to subtract the zero mode piece of the perturbations by imposing $P_0(a_s)=P_0(\mathfrak{u}(\xi)\widetilde{a}_{-s}\mathfrak{u}(\xi)^{\dagger})=0$. As pointed out before, we do not have a definition of the zero mode projector for global perturbations, beyond simply applying the formula \eqref{eq:zeromodeprojector}. For now, we simply assume that we have chosen the operators $\tilde{a}$ and $\mathfrak{b}$ such that there is no nontrivial zero mode piece to the perturbation. The modular Berry curvature associated with the two perturbations is then given by
\beq \label{eq:F(s)}
F(s) = P_0\left(\left[a_{s}, \mathfrak{u}(\xi)\widetilde{a}_{-s}\mathfrak{u}(\xi)^{\dagger}\right]\right)~,
\eeq
where we have made the dependence of the curvature on the modular time $s$ explicit. Generically, the unitary $\mathfrak{u}(\xi)$ does not commute with all operators in $\mathcal{A}$, nor with all operators in $\mathcal{A}'$, and therefore one might expect that there is a nontrivial Berry curvature. We can get rid of  the overall zero mode projector by considering the expectation value of the operator $F(s)$ in the original cyclic and separating state $|\psi\rangle$, which we denote by 
\beq \label{eq:FOTOC}
F_{\psi}(s)\equiv \langle \psi |\left[a_{s}, \mathfrak{u}(\xi)\widetilde{a}_{-s}\mathfrak{u}(\xi)^{\dagger}\right]|\psi \rangle~.
\eeq
The modular time-dependence of the above quantity is determined by the nature of the global unitary that is used to couple the algebra and its commutant. For example, a simple choice of two-sided operator involves combining two operators $b\in \mathcal{A}$ and $\widetilde{b}\in \mathcal{A}'$ into a single global operator $\mathfrak{b}=b\widetilde{b}$. In that case, one can expand the unitary $\mathfrak{u}(\xi)$ in \eqref{eq:u(xi)} up to first order in $\xi$ to find that 
\beq 
\mathfrak{u}(\xi)\widetilde{a}_{-s}\mathfrak{u}(\xi)^{\dagger} = 1+i \xi \left[b\widetilde{b},\widetilde{a}_{-s}\right]+\cdots = 1+i\xi b\left[\widetilde{b},\widetilde{a}_{-s}\right]+\cdots ~,
\eeq
where we have used that $\left[b,\widetilde{a}_{-s}\right]=0$. Therefore, neglecting higher order terms in the coupling parameter $\xi$ the expectation value of the Berry curvature takes the form 
\beq 
F_{\psi}(s) \approx i \xi \langle \psi |\left[a_{s},b\left[\widetilde{b} ,\widetilde{a}_{-s}\right]\right] |\psi \rangle= -i\xi \langle \psi |\left[a_{s},b\right] \left[\widetilde{a}_{-s}, \widetilde{b}\right]|\psi\rangle ~,
\eeq
where we have used that $\left[\widetilde{b} ,\widetilde{a}_{-s}\right]$ commutes with $a_s$ to pull it out of the commutator. Using the identity $J_{\psi}=\Delta_{\psi}^{-1/2}S_{\psi}$, we can equivalently write this as 
\beq \label{eq:squaredcom}
F_{\psi}(s) \approx -i \xi \langle \psi |\left[a_{s},b\right]\Delta_{\psi}^{-1/2} \left[a_{s}, b\right]^{\dagger}|\psi\rangle ~.
\eeq
Expectation values of squared commutators such as the one above have appeared before in the study of quantum chaos \cite{Shenker:2013pqa, Maldacena:2015waa}, and we therefore expect the modular Berry curvature in \eqref{eq:FOTOC} to be relevant to that discussion as well. Indeed, let us go back to equation \eqref{eq:FOTOC} for $F_{\psi}(s)$, and take its Hermitian conjugate. Assuming that the operator $a$ is Hermitian, $a^{\dagger}=a$, we find that 
\beq 
\overline{F_{\psi}(s)} = \langle \psi |[\mathfrak{u}(\xi)\widetilde{a}_{-s}\mathfrak{u}(\xi)^{\dagger},a_{s}] |\psi \rangle= - \langle \psi |[a_{s}, \mathfrak{u}(\xi)\widetilde{a}_{-s}\mathfrak{u}(\xi)^{\dagger}] |\psi \rangle= - F_{\psi}(s)~,
\eeq 
which shows that the quantity $F_{\Psi}(s)$ is purely imaginary. In fact, it is the imaginary part of the following correlator
\beq \label{eq:COTOC}
C_{\psi}(s) 
\equiv 2 i \langle \psi | a_s \mathfrak{u}(\xi) \widetilde{a}_{-s} \mathfrak{u}(\xi)^{\dagger}| \psi \rangle~,
\eeq
as can be readily checked by computing
\beq 
\mathfrak{Im}\, C_{\psi}(s)= \frac{1}{2i}\left( C_{\psi}(s) - \overline{C_{\psi}(s)} \right) = F_{\psi}(s)~.
\eeq
We will refer to the quantity $C_{\psi}(s)$ in terms of a \emph{modular} out-of-time-ordered correlator, as it involves the insertion of operators at non-consecutive modular times.
We would like to point out that the recent paper \cite{Cirafici:2024jdw} introduces a quantity very similar to \eqref{eq:squaredcom} as a probe for modular chaos in the context of operator algebras.

We would now like to argue that the behavior of \eqref{eq:COTOC} as a function of modular time $s$ is a probe for modular chaos in the underlying theory. Let us first assume that $\xi$ is small enough so that we can use the following approximation, where the operator $\mathfrak{u}(\xi)^{\dagger}$ acts by multiplication with a phase on the cyclic and separating state:
\beq 
\mathfrak{u}(\xi)^{\dagger}|\psi\rangle \approx e^{-i \xi \langle \mathfrak{b} \rangle }|\psi\rangle~,
\eeq
In this approximation, the modular OTOC $C_{\psi}$ can be written as
\beq \label{eq:Cpsi1}
C_{\psi}(s) 
\approx e^{-i \xi \langle\mathfrak{b}\rangle} \langle \psi | a_s \mathfrak{u}(\xi) \widetilde{a}_{-s}| \psi \rangle~.
\eeq
Let us now study the $s$-dependence of this expression in more detail. Since we are working in the limit where the coupling $\xi$ between the algebra and its commutant is small, we can expand the expression in \eqref{eq:Cpsi1} as
\beq 
C_{\psi}(s) 
\approx i \xi e^{-i \xi \langle \mathfrak{b}\rangle} \langle \psi | a_s \mathfrak{b} \widetilde{a}_{-s}| \psi \rangle = i \xi e^{-i\xi \langle \mathfrak{b}\rangle} \langle \psi | a \mathfrak{b}_s \widetilde{a}| \psi \rangle~,
\eeq
where at the second equality we used the identity $\Delta_{\psi}|\psi\rangle = |\psi\rangle$ to transfer the modular flow from the operators $a$ and $\widetilde{a}$ to the global operator $\mathfrak{b}$. Therefore, we infer from the above expression that the $s$-dependence of $C_{\psi}$ is determined by the behavior of the modular time-evolved operator $\mathfrak{b}_s$. In the case of a half-sided modular inclusion, the global operator $\mathfrak{b}$ that we have introduced to couple both sides can generically have a piece that is proportional to a modular scrambling mode $G_{+}$. For large $s$, this will give the dominant contribution to $\mathfrak{b}_s$ \cite{DeBoer:2019kdj}. By virtue of the commutation relation in \eqref{eq:commutationG+}, the operator will have an exponential growth of the form 
\beq
\mathfrak{b}_s = \alpha \Delta^{-is}_{\psi}G_{+}\Delta_{\psi}^{is} +\cdots = \alpha e^{2\pi s}G_{+}+\cdots~, 
\eeq
where $\alpha$ is some coefficient and the terms $\cdots$ are subleading in the limit $s\to \infty$. Hence, we find that the presence of a modular scrambling mode leads to an exponentially growing correlator of the form
\beq \label{eq:expgrowth}
C_{\psi}(s) \sim e^{2\pi s} C_{\psi}(0) ~, \qquad \mathrm{as} \qquad s\to \infty~.
\eeq
Physically, the exponential growth of this quantity is a signature of the system's underlying modular chaos. Indeed, if we think about the operator $\mathfrak{b}$ as perturbing the modular Hamiltonian, the growth of these perturbations is exponentially bounded. The modular chaos bound is precisely saturated by the modular scrambling mode $G_+$ \cite{DeBoer:2019kdj}, which shows that the growth according to \eqref{eq:expgrowth} is a sign of maximal chaos. In that sense, the modular time dependence of the Berry curvature \eqref{eq:F(s)} provides a probe for modular chaos in the context of von Neumann algebras.

%% file: discussion.tex
\section{Discussion}
In the present paper, we have studied the modular Berry phase from an operator algebraic perspective. Using techniques from the modular theory of von Neumann algebras, we have set up a formalism that associates a geometric phase to a one-parameter family of modular operators. The upshot of this more abstract approach is that it does not rely on any assumptions about the type of von Neumann algebra, and it allows one to define the modular Berry phase even when the usual notions of pure state or density matrix do not make sense. While this resolves some technical subtleties that appeared in previous works on the modular Berry phase, the algebraic approach, as we have argued, also reveals some new physics, specifically in the case of type III$_1$ von Neumann algebras.

This is particularly relevant for the AdS/CFT correspondence, where subregions in the spacetime geometry emerge from boundary algebras that are characterized as type III$_1$ von Neumann algebras in the large $N$ limit. It has been argued that some emergent properties, like causality, spacetime symmetries and the concept of a horizon, can be explained from the type III nature of the boundary algebra. One of the main results of the present work, is the discovery of a novel emergent quantity, namely an emergent symplectic form in the large $N$ limit. The symplectic form arises as a particular matrix element of the modular Berry curvature associated to a class a state-deformations in the type III$_1$ boundary algebra. This computation relies mostly on the generalized free field description of the boundary theory in the strict $N\to\infty$ limit. In addition, we have shown that for a different type of deformation, generated by modular scrambling modes, the modular Berry curvature can be used to probe another aspect of the local structure of spacetime, namely the existence of an emergent Poincar\'e algebra in the bulk, and we highlight its relation to some underlying modular chaos in the boundary algebra.   

There remain a few subtleties that need to be addressed. An important distinction that we have a made in our treatment of the modular Berry phase is the difference between \emph{state-changing} and \emph{shape-changing} deformations. In the former case, we consider perturbations that change the global state and map the algebra to itself via an automorphism (``two states, one algebra''), while in the latter case the algebra is deformed and the state kept fixed (``two algebras, one state''). As a representative example of the shape-changing case, we have looked at modular scrambling modes which arise as positive generators that can be used to ``translate'' the algebra. Those operators exist when we have the algebraic structure of a half-sided modular inclusion. One important observation is that the shape-changing transformations cannot be part of the original algebra. Instead, they are properly implemented as global transformations. To define a modular Berry phase for global transformations turns out to be subtle, and in Section \ref{sec:modchaos} we have not been very explicit about the subalgebra of \emph{global} modular zero modes, i.e., the modular conserved charges, or the extension of the zero mode projector to the space of global transformations. We have argued that in special cases, for example, when the space of deformations has the structure of a finite-dimensional Lie algebra (like in the case of the conformal group), one can use an inner product to define a zero mode projection (see, for example, \cite{Czech:2023zmq}). We have, however, not found a satisfying resolution of this issue that applies in general situations. This deserves further thought.   

Moreover, it would be interesting to see how the results in Section \ref{sec:symform} are modified in the presence of $1/N$ corrections. One unusual aspect of the GFF computation is that the resulting modular Berry curvature is proportional to the identity operator $\mathds{1}$. The modular operator $\Delta_{\Psi}$ does not appear in the final expression, because a priori it is not an element of the zero mode algebra $\mathcal{A}^{\psi}$. We expect that including $1/N$ corrections in the above calculation corresponds, mathematically, to taking the crossed product of the algebra with modular automorphism group \cite{Witten:2021unn}. Starting from the type III$_1$ GFF algebra, the resulting modular crossed product algebra is a type II$_{\infty}$ von Neumann algebra and its zero mode subalgebra gets enhanced by the addition of a new generator that implements the modular flow. One can imagine that a similar construction can be used to add additional modular zero modes. Given a group $H$ of modular zero modes that are realized as automorphisms of the algebra, the correct algebraic structure that includes those zero modes is the crossed product algebra $\mathcal{A}\rtimes H$ (see, e.g., \cite{Fewster:2024pur, AliAhmad:2024eun} for a discussion of the crossed product in the presence of more general symmetry groups). It would be interesting to understand whether there is still some relation to a symplectic form in the case of additional zero modes. A compelling hint towards such an interpretation might be found in \cite{Klinger:2023tgi,Klinger:2023auu}, where a connection between the crossed product construction and the structure of an extended phase space is revealed. We expect that the modular Berry curvature, which takes values in the Lie algebra of $H$, agrees with the symplectic form on this extended phase space. It is worth checking this in detail.

As a final comment, we would like to discuss the connection between the modular Berry phase and quantum chaos. It has  been recently argued the relevant algebraic properties can be organized in terms of a quantum ergodic hierarchy \cite{Furuya:2023fei, Gesteau:2023rrx, Ouseph:2023juq}. We have shown that the decay of the modular Berry curvature for a certain class of shape-changing perturbations signals the emergence of a local Poincar\'e algebra, which fits nicely into this ergodic hierarchy. In addition, we have defined a modular version of the OTOC in terms of geometric phase that exhibits maximal exponential growth in the presence of modular scrambling modes. On the gravity side, OTOC physics is captured by high energy scattering in the near-horizon region of a black hole, which is well-described by a semi-classical shockwave geometry. It would be interesting to see if the relation between the Berry curvature and the bulk symplectic form holds in this setting (where we are considering global perturbations) as well. A first step would be computing the symplectic form in such a shockwave background (see, e.g., \cite{Liu:2021kay}). On a slightly different note, the Berry phase is in certain cases expected to be related to complexity, especially in the context of quantum information geometry \cite{Belin:2018bpg}. It would be interesting to investigate whether our algebraic framework might provide a first step towards a satisfying von Neumann algebra version of complexity (one such a proposal has been recently explored in \cite{Hollands:2023abp}).

%% file: appendix.tex
\section{A global zero mode computation}

\label{app:A}

In this appendix, we have included a computation of the zero mode by applying the integral formula \eqref{eq:zeromodeprojector} to the global operator $\mathfrak{u}_+(t)\in B(\mathcal{H})$. We have to evaluate the following integral: 
\beq 
\mathcal{E}_0(\mathfrak{u}_+(t))= \lim_{\Lambda \to \infty}\frac{1}{2\Lambda}\int_{-\Lambda}^{\Lambda}ds\,\Delta_\psi^{-is}\mathfrak{u}_+(t)\Delta_\psi^{is}=\lim_{\Lambda \to \infty}\frac{1}{2\Lambda}\int_{-\Lambda}^{\Lambda}ds\, \mathfrak{u}_{+}(e^{2\pi s}t)~.
\eeq
Since the integrand in the above expression is a bounded operator we expect the integral to make sense for finite $\Lambda$. However, it is not clear that the expression has a good $\Lambda \to \infty$ limit. 

To this end, we consider the following integral for $a>0$,
\beq \label{eq:expiexp}
\frac{1}{2\Lambda}\int_{-\Lambda}^{\Lambda}ds\,\exp\left(ie^{2\pi s}t a\right)=\frac{1}{2\Lambda}\int_{e^{-2\pi\Lambda}}^{e^{2\pi\Lambda}}\frac{dr}{2\pi r}\,\exp(ir t a)~,
\eeq
where we have introduced the change of coordinate $s=\frac{1}{2\pi}\log r$. The right-hand side of \eqref{eq:expiexp} can be expressed in terms of special functions as 
\beq \label{eq:integralEi}
\frac{1}{2\Lambda}\int_{e^{-2\pi\Lambda}}^{e^{2\pi\Lambda}}\frac{dr}{2\pi r}\,\exp(ir t a) = \frac{1}{2\Lambda}\left( \mathrm{Ei}(i e^{2\pi \Lambda} ta)- \mathrm{Ei}(ie^{-2\pi \Lambda} ta)\right)~,
\eeq
where $\mathrm{Ei}(x)$ is the so-called exponential integral. For $x\in \mathbb{R}\setminus \{0\}$, the exponential integral function is defined via the integral formula (see, for example, \cite{Abramowitz})
\beq 
\mathrm{Ei}(x)\equiv \int^{x}_{-\infty} \frac{e^{t}}{t} \,dt~.
\eeq
Its extension to purely imaginary values $ix$ with $x>0$ can be expressed in terms of the sine and cosine integrals
\beq 
\mathrm{Ci}(x) \equiv - \int^{\infty}_{x} \frac{\cos{t}}{t} \,dt~, \qquad \mathrm{Si}(x) \equiv \int^{x}_{0} \frac{\sin{t} }{t}\,dt~,
\eeq
via the identity
\beq 
\mathrm{Ei}(ix) = - \frac{i \pi}{2} + \mathrm{Ci}(x)+i \mathrm{Si}(x)~.
\eeq
To compute the $\Lambda \to \infty$ limit of this expression, we use the following series expansions,
\beq \label{eq:series1}
\mathrm{Ei}(ix)\sim \pi i + \frac{e^{ix}}{x}+\cdots~, \qquad x \to \infty~,
\eeq
and 
\beq \label{eq:series2}
\mathrm{Ei}(ix)\sim -\frac{\pi i}{2} + \gamma + \ln x + \cdots ~, \qquad x \to 0~,
\eeq
where $\gamma$ is the Euler-Mascheroni constant. Setting $x=e^{2\pi \Lambda} ta$ in \eqref{eq:series1} and $x=e^{-2\pi \Lambda} ta$ in \eqref{eq:series2}, we find that for large $\Lambda$ the integral in \eqref{eq:integralEi} can be expanded to give 
\beq
\frac{1}{2\Lambda}\int_{e^{-2\pi\Lambda}}^{e^{2\pi\Lambda}}\frac{dr}{2\pi r}\,\exp(ir t a) = \frac{1}{2\Lambda}\left(\frac{3\pi i}{2}-\gamma + 2\pi \Lambda -\ln(ta)+\mathcal{O}(e^{-2\pi \Lambda})\right)~.
\eeq
In taking the $\Lambda \to \infty$ limit, most terms drop out and the final result is given by
\beq 
\lim_{\Lambda \to \infty}\frac{1}{2\Lambda}\int_{-\Lambda}^{\Lambda}ds\,\exp\left(ie^{2\pi s}t a\right) = \pi~.
\eeq
The fact that the right-hand side does not depend on the parameter $a>0$ suggests that $\mathcal{E}_0(\mathfrak{u}_{+}(t))=\pi \mathds{1}$ for all $t>0$. Using \eqref{eq:P0definition}, we are led to the conclusion that $P_{0}(G_{+})=0$, as expected. More generally, this computation shows that an application of \eqref{eq:zeromodeprojector} to global transformations does exhibit the behavior that one would expect from a bona fide modular zero mode projector, in the sense that the corresponding $P_0$ removes eigenoperators with respect to the adjoint action of the modular Hamiltonian with non-zero eigenvalue.

%% file: paper.bbl
\providecommand{\href}[2]{#2}\begingroup\raggedright\begin{thebibliography}{10}

\bibitem{DeBoer:2019kdj}
J.~De~Boer and L.~Lamprou, \emph{{Holographic Order from Modular Chaos}},
  \href{https://doi.org/10.1007/JHEP06(2020)024}{\emph{JHEP} {\bfseries 06}
  (2020) 024} [\href{https://arxiv.org/abs/1912.02810}{{\ttfamily
  1912.02810}}].

\bibitem{Kang:2018xqy}
M.J.~Kang and D.K.~Kolchmeyer, \emph{{Holographic Relative Entropy in
  Infinite-Dimensional Hilbert Spaces}},
  \href{https://doi.org/10.1007/s00220-022-04627-z}{\emph{Commun. Math. Phys.}
  {\bfseries 400} (2023) 1665}
  [\href{https://arxiv.org/abs/1811.05482}{{\ttfamily 1811.05482}}].

\bibitem{Gesteau:2020rtg}
E.~Gesteau and M.J.~Kang, \emph{{Thermal states are vital: Entanglement Wedge
  Reconstruction from Operator-Pushing}},
  \href{https://arxiv.org/abs/2005.07189}{{\ttfamily 2005.07189}}.

\bibitem{Leutheusser:2021qhd}
S.~Leutheusser and H.~Liu, \emph{{Causal connectability between quantum systems
  and the black hole interior in holographic duality}},
  \href{https://doi.org/10.1103/PhysRevD.108.086019}{\emph{Phys. Rev. D}
  {\bfseries 108} (2023) 086019}
  [\href{https://arxiv.org/abs/2110.05497}{{\ttfamily 2110.05497}}].

\bibitem{Leutheusser:2021frk}
S.~Leutheusser and H.~Liu, \emph{{Emergent times in holographic duality}},
  \href{https://doi.org/10.1103/PhysRevD.108.086020}{\emph{Phys. Rev. D}
  {\bfseries 108} (2023) 086020}
  [\href{https://arxiv.org/abs/2112.12156}{{\ttfamily 2112.12156}}].

\bibitem{Chandrasekaran:2022eqq}
V.~Chandrasekaran, G.~Penington and E.~Witten, \emph{{Large N algebras and
  generalized entropy}},
  \href{https://doi.org/10.1007/JHEP04(2023)009}{\emph{JHEP} {\bfseries 04}
  (2023) 009} [\href{https://arxiv.org/abs/2209.10454}{{\ttfamily
  2209.10454}}].

\bibitem{Leutheusser:2022bgi}
S.~Leutheusser and H.~Liu, \emph{{Subregion-subalgebra duality: Emergence of
  space and time in holography}},
  \href{https://doi.org/10.1103/PhysRevD.111.066021}{\emph{Phys. Rev. D}
  {\bfseries 111} (2025) 066021}
  [\href{https://arxiv.org/abs/2212.13266}{{\ttfamily 2212.13266}}].

\bibitem{deBoer:2022zps}
J.~de~Boer, D.L.~Jafferis and L.~Lamprou, \emph{{On black hole interior
  reconstruction, singularities and the emergence of time}},
  \href{https://arxiv.org/abs/2211.16512}{{\ttfamily 2211.16512}}.

\bibitem{Furuya:2023fei}
K.~Furuya, N.~Lashkari, M.~Moosa and S.~Ouseph, \emph{{Information loss, mixing
  and emergent type III$_{1}$ factors}},
  \href{https://doi.org/10.1007/JHEP08(2023)111}{\emph{JHEP} {\bfseries 08}
  (2023) 111} [\href{https://arxiv.org/abs/2305.16028}{{\ttfamily
  2305.16028}}].

\bibitem{Gesteau:2023rrx}
E.~Gesteau, \emph{{Emergent spacetime and the ergodic hierarchy}},
  \href{https://doi.org/10.1103/PhysRevD.110.106005}{\emph{Phys. Rev. D}
  {\bfseries 110} (2024) 106005}
  [\href{https://arxiv.org/abs/2310.13733}{{\ttfamily 2310.13733}}].

\bibitem{Ouseph:2023juq}
S.~Ouseph, K.~Furuya, N.~Lashkari, K.L.~Leung and M.~Moosa, \emph{{Local
  Poincar\'e algebra from quantum chaos}},
  \href{https://doi.org/10.1007/JHEP01(2024)112}{\emph{JHEP} {\bfseries 01}
  (2024) 112} [\href{https://arxiv.org/abs/2310.13736}{{\ttfamily
  2310.13736}}].

\bibitem{Witten:2023xze}
E.~Witten, \emph{{A background-independent algebra in quantum gravity}},
  \href{https://doi.org/10.1007/JHEP03(2024)077}{\emph{JHEP} {\bfseries 03}
  (2024) 077} [\href{https://arxiv.org/abs/2308.03663}{{\ttfamily
  2308.03663}}].

\bibitem{engelhardt:2023xer}
N.~Engelhardt and H.~Liu, \emph{{Algebraic ER=EPR and complexity transfer}},
  \href{https://doi.org/10.1007/JHEP07(2024)013}{\emph{JHEP} {\bfseries 07}
  (2024) 013} [\href{https://arxiv.org/abs/2311.04281}{{\ttfamily
  2311.04281}}].

\bibitem{Gesteau:2024dhj}
E.~Gesteau and L.~Santilli, \emph{{Explicit large $N$ von Neumann algebras from
  matrix models}}, \href{https://doi.org/10.4310/ATMP.241031230051}{\emph{Adv.
  Theor. Math. Phys.} {\bfseries 28} (2024) 2245}
  [\href{https://arxiv.org/abs/2402.10262}{{\ttfamily 2402.10262}}].

\bibitem{vanderHeijden:2024tdk}
J.~van~der Heijden and E.~Verlinde, \emph{{An operator algebraic approach to
  black hole information}},
  \href{https://doi.org/10.1007/JHEP02(2025)207}{\emph{JHEP} {\bfseries 02}
  (2025) 207} [\href{https://arxiv.org/abs/2408.00071}{{\ttfamily
  2408.00071}}].

\bibitem{Gesteau:2024rpt}
E.~Gesteau and H.~Liu, \emph{{Toward stringy horizons}},
  \href{https://arxiv.org/abs/2408.12642}{{\ttfamily 2408.12642}}.

\bibitem{Lashkari:2024lkt}
N.~Lashkari, K.L.~Leung, M.~Moosa and S.~Ouseph, \emph{{Modular Intersections,
  Time Interval Algebras and Stringy AdS$_2$}},
  \href{https://arxiv.org/abs/2412.19882}{{\ttfamily 2412.19882}}.

\bibitem{Lashkari:2018oke}
N.~Lashkari, H.~Liu and S.~Rajagopal, \emph{{Modular flow of excited states}},
  \href{https://doi.org/10.1007/JHEP09(2021)166}{\emph{JHEP} {\bfseries 09}
  (2021) 166} [\href{https://arxiv.org/abs/1811.05052}{{\ttfamily
  1811.05052}}].

\bibitem{Gesteau:2021jzp}
E.~Gesteau and M.J.~Kang, \emph{{Nonperturbative gravity corrections to bulk
  reconstruction}}, \href{https://doi.org/10.1088/1751-8121/acef7d}{\emph{J.
  Phys. A} {\bfseries 56} (2023) 385401}
  [\href{https://arxiv.org/abs/2112.12789}{{\ttfamily 2112.12789}}].

\bibitem{Crann:2024gkv}
J.~Crann and M.J.~Kang, \emph{{Algebraic approach to spacetime bulk
  reconstruction}},  \href{https://arxiv.org/abs/2412.00298}{{\ttfamily
  2412.00298}}.

\bibitem{Chandrasekaran:2022cip}
V.~Chandrasekaran, R.~Longo, G.~Penington and E.~Witten, \emph{{An algebra of
  observables for de Sitter space}},
  \href{https://doi.org/10.1007/JHEP02(2023)082}{\emph{JHEP} {\bfseries 02}
  (2023) 082} [\href{https://arxiv.org/abs/2206.10780}{{\ttfamily
  2206.10780}}].

\bibitem{Penington:2023dql}
G.~Penington and E.~Witten, \emph{{Algebras and States in JT Gravity}},
  \href{https://arxiv.org/abs/2301.07257}{{\ttfamily 2301.07257}}.

\bibitem{Kolchmeyer:2023gwa}
D.K.~Kolchmeyer, \emph{{von Neumann algebras in JT gravity}},
  \href{https://doi.org/10.1007/JHEP06(2023)067}{\emph{JHEP} {\bfseries 06}
  (2023) 067} [\href{https://arxiv.org/abs/2303.04701}{{\ttfamily
  2303.04701}}].

\bibitem{Jensen:2023yxy}
K.~Jensen, J.~Sorce and A.J.~Speranza, \emph{{Generalized entropy for general
  subregions in quantum gravity}},
  \href{https://doi.org/10.1007/JHEP12(2023)020}{\emph{JHEP} {\bfseries 12}
  (2023) 020} [\href{https://arxiv.org/abs/2306.01837}{{\ttfamily
  2306.01837}}].

\bibitem{AliAhmad:2023etg}
S.~Ali~Ahmad and R.~Jefferson, \emph{{Crossed product algebras and generalized
  entropy for subregions}},
  \href{https://doi.org/10.21468/SciPostPhysCore.7.2.020}{\emph{SciPost Phys.
  Core} {\bfseries 7} (2024) 020}
  [\href{https://arxiv.org/abs/2306.07323}{{\ttfamily 2306.07323}}].

\bibitem{Kudler-Flam:2023qfl}
J.~Kudler-Flam, S.~Leutheusser and G.~Satishchandran, \emph{{Generalized black
  hole entropy is von Neumann entropy}},
  \href{https://doi.org/10.1103/PhysRevD.111.025013}{\emph{Phys. Rev. D}
  {\bfseries 111} (2025) 025013}
  [\href{https://arxiv.org/abs/2309.15897}{{\ttfamily 2309.15897}}].

\bibitem{Kudler-Flam:2023hkl}
J.~Kudler-Flam, S.~Leutheusser, A.A.~Rahman, G.~Satishchandran and
  A.J.~Speranza, \emph{{Covariant regulator for entanglement entropy: Proofs of
  the Bekenstein bound and the quantum null energy condition}},
  \href{https://doi.org/10.1103/PhysRevD.111.105001}{\emph{Phys. Rev. D}
  {\bfseries 111} (2025) 105001}
  [\href{https://arxiv.org/abs/2312.07646}{{\ttfamily 2312.07646}}].

\bibitem{Faulkner:2024gst}
T.~Faulkner and A.J.~Speranza, \emph{{Gravitational algebras and the
  generalized second law}},
  \href{https://doi.org/10.1007/JHEP11(2024)099}{\emph{JHEP} {\bfseries 11}
  (2024) 099} [\href{https://arxiv.org/abs/2405.00847}{{\ttfamily
  2405.00847}}].

\bibitem{Chen:2024rpx}
C.-H.~Chen and G.~Penington, \emph{{A clock is just a way to tell the time:
  gravitational algebras in cosmological spacetimes}},
  \href{https://arxiv.org/abs/2406.02116}{{\ttfamily 2406.02116}}.

\bibitem{Kudler-Flam:2024psh}
J.~Kudler-Flam, S.~Leutheusser and G.~Satishchandran, \emph{{Algebraic
  Observational Cosmology}},
  \href{https://arxiv.org/abs/2406.01669}{{\ttfamily 2406.01669}}.

\bibitem{Kolchmeyer:2024fly}
D.K.~Kolchmeyer and H.~Liu, \emph{{Chaos and the Emergence of the Cosmological
  Horizon}},  \href{https://arxiv.org/abs/2411.08090}{{\ttfamily 2411.08090}}.

\bibitem{Jensen:2024dnl}
K.~Jensen, S.~Raju and A.J.~Speranza, \emph{{Holographic observers for
  time-band algebras}},
  \href{https://doi.org/10.1007/JHEP06(2025)242}{\emph{JHEP} {\bfseries 06}
  (2025) 242} [\href{https://arxiv.org/abs/2412.21185}{{\ttfamily
  2412.21185}}].

\bibitem{AliAhmad:2025ukh}
S.~Ali~Ahmad and R.~Jefferson, \emph{{Algebraic perturbation theory:
  traversable wormholes and generalized entropy beyond subleading order}},
  \href{https://arxiv.org/abs/2501.01487}{{\ttfamily 2501.01487}}.

\bibitem{Chemissany:2025vye}
W.~Chemissany, E.~Gesteau, A.~Jahn, D.~Murphy and L.~Shaposhnik, \emph{{On
  Infinite Tensor Networks, Complementary Recovery and Type II Factors}},
  \href{https://arxiv.org/abs/2504.00096}{{\ttfamily 2504.00096}}.

\bibitem{Speranza:2025joj}
A.J.~Speranza, \emph{{An intrinsic cosmological observer}},
  \href{https://arxiv.org/abs/2504.07630}{{\ttfamily 2504.07630}}.

\bibitem{Berry:1984jv}
M.V.~Berry, \emph{{Quantal phase factors accompanying adiabatic changes}},
  \href{https://doi.org/10.1098/rspa.1984.0023}{\emph{Proc. Roy. Soc. Lond. A}
  {\bfseries 392} (1984) 45}.

\bibitem{Czech:2017zfq}
B.~Czech, L.~Lamprou, S.~Mccandlish and J.~Sully, \emph{{Modular Berry
  Connection for Entangled Subregions in AdS/CFT}},
  \href{https://doi.org/10.1103/PhysRevLett.120.091601}{\emph{Phys. Rev. Lett.}
  {\bfseries 120} (2018) 091601}
  [\href{https://arxiv.org/abs/1712.07123}{{\ttfamily 1712.07123}}].

\bibitem{Czech:2019vih}
B.~Czech, J.~De~Boer, D.~Ge and L.~Lamprou, \emph{{A Modular Sewing Kit for
  Entanglement Wedges}},
  \href{https://doi.org/10.1007/JHEP11(2019)094}{\emph{JHEP} {\bfseries 11}
  (2019) 094} [\href{https://arxiv.org/abs/1903.04493}{{\ttfamily
  1903.04493}}].

\bibitem{Czech:2015qta}
B.~Czech, L.~Lamprou, S.~McCandlish and J.~Sully, \emph{{Integral Geometry and
  Holography}}, \href{https://doi.org/10.1007/JHEP10(2015)175}{\emph{JHEP}
  {\bfseries 10} (2015) 175}
  [\href{https://arxiv.org/abs/1505.05515}{{\ttfamily 1505.05515}}].

\bibitem{Czech:2016xec}
B.~Czech, L.~Lamprou, S.~McCandlish, B.~Mosk and J.~Sully, \emph{{A
  Stereoscopic Look into the Bulk}},
  \href{https://doi.org/10.1007/JHEP07(2016)129}{\emph{JHEP} {\bfseries 07}
  (2016) 129} [\href{https://arxiv.org/abs/1604.03110}{{\ttfamily
  1604.03110}}].

\bibitem{deBoer:2015kda}
J.~de~Boer, M.P.~Heller, R.C.~Myers and Y.~Neiman, \emph{{Holographic de Sitter
  Geometry from Entanglement in Conformal Field Theory}},
  \href{https://doi.org/10.1103/PhysRevLett.116.061602}{\emph{Phys. Rev. Lett.}
  {\bfseries 116} (2016) 061602}
  [\href{https://arxiv.org/abs/1509.00113}{{\ttfamily 1509.00113}}].

\bibitem{deBoer:2016pqk}
J.~de~Boer, F.M.~Haehl, M.P.~Heller and R.C.~Myers, \emph{{Entanglement,
  Holography and Causal Diamonds}},
  \href{https://doi.org/10.1007/JHEP08(2016)162}{\emph{JHEP} {\bfseries 08}
  (2016) 162} [\href{https://arxiv.org/abs/1606.03307}{{\ttfamily
  1606.03307}}].

\bibitem{Chen:2022nwf}
B.~Chen, B.~Czech, L.-Y.~Hung and G.~Wong, \emph{{Modular Parallel Transport of
  Multiple Intervals in 1+1-Dimensional Free Fermion Theory}},
  \href{https://doi.org/10.1007/JHEP03(2023)147}{\emph{JHEP} {\bfseries 03}
  (2023) 147} [\href{https://arxiv.org/abs/2211.12545}{{\ttfamily
  2211.12545}}].

\bibitem{Aalsma:2024qnf}
L.~Aalsma, C.~Keeler and C.~Zukowski, \emph{{Quantum extremal modular
  curvature: modular transport with islands}},
  \href{https://doi.org/10.1007/JHEP10(2024)006}{\emph{JHEP} {\bfseries 10}
  (2024) 006} [\href{https://arxiv.org/abs/2406.05176}{{\ttfamily
  2406.05176}}].

\bibitem{deBoer:2021zlm}
J.~de~Boer, R.~Espindola, B.~Najian, D.~Patramanis, J.~van~der Heijden and
  C.~Zukowski, \emph{{Virasoro Entanglement Berry Phases}},
  \href{https://doi.org/10.1007/JHEP03(2022)179}{\emph{JHEP} {\bfseries 03}
  (2022) 179} [\href{https://arxiv.org/abs/2111.05345}{{\ttfamily
  2111.05345}}].

\bibitem{Czech:2023zmq}
B.~Czech, J.~de~Boer, R.~Espindola, B.~Najian, J.~van~der Heijden and
  C.~Zukowski, \emph{{Changing states in holography: From modular Berry
  curvature to the bulk symplectic form}},
  \href{https://doi.org/10.1103/PhysRevD.108.066003}{\emph{Phys. Rev. D}
  {\bfseries 108} (2023) 066003}
  [\href{https://arxiv.org/abs/2305.16384}{{\ttfamily 2305.16384}}].

\bibitem{Banerjee:2023eew}
S.~Banerjee, M.~Dorband, J.~Erdmenger and A.-L.~Weigel, \emph{{Geometric phases
  characterise operator algebras and missing information}},
  \href{https://doi.org/10.1007/JHEP10(2023)026}{\emph{JHEP} {\bfseries 10}
  (2023) 026} [\href{https://arxiv.org/abs/2306.00055}{{\ttfamily
  2306.00055}}].

\bibitem{Nogueira:2021ngh}
F.S.~Nogueira, S.~Banerjee, M.~Dorband, R.~Meyer, J.v.d.~Brink and
  J.~Erdmenger, \emph{{Geometric phases distinguish entangled states in
  wormhole quantum mechanics}},
  \href{https://doi.org/10.1103/PhysRevD.105.L081903}{\emph{Phys. Rev. D}
  {\bfseries 105} (2022) L081903}
  [\href{https://arxiv.org/abs/2109.06190}{{\ttfamily 2109.06190}}].

\bibitem{Banerjee:2022jnv}
S.~Banerjee, M.~Dorband, J.~Erdmenger, R.~Meyer and A.-L.~Weigel, \emph{{Berry
  phases, wormholes and factorization in AdS/CFT}},
  \href{https://doi.org/10.1007/JHEP08(2022)162}{\emph{JHEP} {\bfseries 08}
  (2022) 162} [\href{https://arxiv.org/abs/2202.11717}{{\ttfamily
  2202.11717}}].

\bibitem{Witten:2018zxz}
E.~Witten, \emph{{APS Medal for Exceptional Achievement in Research: Invited
  Article on Entanglement Properties of Quantum Field Theory}},
  \href{https://doi.org/10.1103/RevModPhys.90.045003}{\emph{Rev. Mod. Phys.}
  {\bfseries 90} (2018) 045003}
  [\href{https://arxiv.org/abs/1803.04993}{{\ttfamily 1803.04993}}].

\bibitem{Sorce:2023fdx}
J.~Sorce, \emph{{Notes on the type classification of von Neumann algebras}},
  \href{https://doi.org/10.1142/S0129055X24300024}{\emph{Rev. Math. Phys.}
  {\bfseries 36} (2024) 2430002}
  [\href{https://arxiv.org/abs/2302.01958}{{\ttfamily 2302.01958}}].

\bibitem{Greenberg:1961mr}
O.W.~Greenberg, \emph{{Generalized Free Fields and Models of Local Field
  Theory}}, \href{https://doi.org/10.1016/0003-4916(61)90032-X}{\emph{Annals
  Phys.} {\bfseries 16} (1961) 158}.

\bibitem{Aharony:1999ti}
O.~Aharony, S.S.~Gubser, J.M.~Maldacena, H.~Ooguri and Y.~Oz, \emph{{Large N
  field theories, string theory and gravity}},
  \href{https://doi.org/10.1016/S0370-1573(99)00083-6}{\emph{Phys. Rept.}
  {\bfseries 323} (2000) 183}
  [\href{https://arxiv.org/abs/hep-th/9905111}{{\ttfamily hep-th/9905111}}].

\bibitem{Duetsch:2002hc}
M.~Duetsch and K.-H.~Rehren, \emph{{Generalized free fields and the AdS - CFT
  correspondence}},
  \href{https://doi.org/10.1007/s00023-003-0141-9}{\emph{Annales Henri
  Poincare} {\bfseries 4} (2003) 613}
  [\href{https://arxiv.org/abs/math-ph/0209035}{{\ttfamily math-ph/0209035}}].

\bibitem{Belin:2018fxe}
A.~Belin, A.~Lewkowycz and G.~S\'arosi, \emph{{The Boundary Dual of the Bulk
  Symplectic Form}},
  \href{https://doi.org/10.1016/j.physletb.2018.10.071}{\emph{Phys. Lett. B}
  {\bfseries 789} (2019) 71}
  [\href{https://arxiv.org/abs/1806.10144}{{\ttfamily 1806.10144}}].

\bibitem{Belin:2018bpg}
A.~Belin, A.~Lewkowycz and G.~S\'arosi, \emph{{Complexity and the Bulk Volume,
  a New York Time Story}},
  \href{https://doi.org/10.1007/JHEP03(2019)044}{\emph{JHEP} {\bfseries 03}
  (2019) 044} [\href{https://arxiv.org/abs/1811.03097}{{\ttfamily
  1811.03097}}].

\bibitem{Kirklin:2019ror}
J.~Kirklin, \emph{{The Holographic Dual of the Entanglement Wedge Symplectic
  Form}}, \href{https://doi.org/10.1007/JHEP01(2020)071}{\emph{JHEP} {\bfseries
  01} (2020) 071} [\href{https://arxiv.org/abs/1910.00457}{{\ttfamily
  1910.00457}}].

\bibitem{Borchers:1991xk}
H.J.~Borchers, \emph{{The CPT theorem in two-dimensional theories of local
  observables}}, \href{https://doi.org/10.1007/BF02099011}{\emph{Commun. Math.
  Phys.} {\bfseries 143} (1992) 315}.

\bibitem{Wiesbrock:1992mg}
H.W.~Wiesbrock, \emph{{Half sided modular inclusions of von Neumann algebras}},
  \href{https://doi.org/10.1007/BF02098019}{\emph{Commun. Math. Phys.}
  {\bfseries 157} (1993) 83}.

\bibitem{Borchers:1995zg}
H.J.~Borchers, \emph{{On the use of modular groups in quantum field theory}},
  {\emph{Ann. Inst. H. Poincare Phys. Theor.} {\bfseries 63} (1995) 331}.

\bibitem{Araki-Zsido}
H.~{Araki} and L.~{Zsid{\'o}}, \emph{{Extension of the Structure Theorem of
  Borchers and its Application to Half-Sided Modular Inclusions}},
  \href{https://doi.org/10.1142/S0129055X05002388}{\emph{Reviews in
  Mathematical Physics} {\bfseries 17} (2005) 491}
  [\href{https://arxiv.org/abs/math/0412061}{{\ttfamily math/0412061}}].

\bibitem{Takesaki:1970aki}
M.~Takesaki, \emph{{Tomita's Theory of Modular Hilbert Algebras and its
  Applications}}, Lecture Notes in Mathematics, Springer-Verlag (1970),
  \href{https://doi.org/10.1007/bfb0065832}{10.1007/bfb0065832}.

\bibitem{takesakiII}
M.~Takesaki, \emph{Theory of Operator Algebras II}, vol.~127 (01, 2003),
  \href{https://doi.org/10.1007/978-1-4612-6188-9}{10.1007/978-1-4612-6188-9}.

\bibitem{Summers:2003tf}
S.J.~Summers, \emph{{Tomita-Takesaki modular theory}},
  \href{https://arxiv.org/abs/math-ph/0511034}{{\ttfamily math-ph/0511034}}.

\bibitem{Longo:1979dw}
R.~Longo, \emph{Notes on algebraic invariants for noncommutative dynamical
  systems}, \href{https://doi.org/10.1007/BF01197443}{\emph{Commun. Math.
  Phys.} {\bfseries 69} (1979) 195}.

\bibitem{2023arXiv230514217M}
A.~{Marrakchi} and S.~{Vaes}, \emph{{Ergodic states on type III$_1$ factors and
  ergodic actions}},
  \href{https://doi.org/10.48550/arXiv.2305.14217}{\emph{arXiv e-prints} (2023)
  arXiv:2305.14217} [\href{https://arxiv.org/abs/2305.14217}{{\ttfamily
  2305.14217}}].

\bibitem{Witten:2021unn}
E.~Witten, \emph{{Gravity and the crossed product}},
  \href{https://doi.org/10.1007/JHEP10(2022)008}{\emph{JHEP} {\bfseries 10}
  (2022) 008} [\href{https://arxiv.org/abs/2112.12828}{{\ttfamily
  2112.12828}}].

\bibitem{AliAhmad:2024eun}
S.~Ali~Ahmad, M.S.~Klinger and S.~Lin, \emph{{Semifinite von Neumann algebras
  in gauge theory and gravity}},
  \href{https://doi.org/10.1103/PhysRevD.111.045006}{\emph{Phys. Rev. D}
  {\bfseries 111} (2025) 045006}
  [\href{https://arxiv.org/abs/2407.01695}{{\ttfamily 2407.01695}}].

\bibitem{DeVuyst:2024pop}
J.~De~Vuyst, S.~Eccles, P.A.~Hoehn and J.~Kirklin, \emph{{Gravitational entropy
  is observer-dependent}},  \href{https://arxiv.org/abs/2405.00114}{{\ttfamily
  2405.00114}}.

\bibitem{takesaki1972conditional}
M.~Takesaki, \emph{Conditional expectations in von neumann algebras},
  {\emph{Journal of Functional Analysis} {\bfseries 9} (1972) 306}.

\bibitem{Sorce:2023gio}
J.~Sorce, \emph{{An intuitive construction of modular flow}},
  \href{https://doi.org/10.1007/JHEP12(2023)079}{\emph{JHEP} {\bfseries 12}
  (2023) 079} [\href{https://arxiv.org/abs/2309.16766}{{\ttfamily
  2309.16766}}].

\bibitem{Faulkner:2020hzi}
T.~Faulkner, \emph{{The holographic map as a conditional expectation}},
  \href{https://arxiv.org/abs/2008.04810}{{\ttfamily 2008.04810}}.

\bibitem{Furuya:2020tzv}
K.~Furuya, N.~Lashkari and S.~Ouseph, \emph{{Real-space RG, error correction
  and Petz map}}, \href{https://doi.org/10.1007/JHEP01(2022)170}{\emph{JHEP}
  {\bfseries 01} (2022) 170}
  [\href{https://arxiv.org/abs/2012.14001}{{\ttfamily 2012.14001}}].

\bibitem{AliAhmad:2024saq}
S.~Ali~Ahmad and M.S.~Klinger, \emph{{Emergent geometry from quantum
  probability}}, \href{https://doi.org/10.1103/PhysRevD.111.105015}{\emph{Phys.
  Rev. D} {\bfseries 111} (2025) 105015}
  [\href{https://arxiv.org/abs/2411.07288}{{\ttfamily 2411.07288}}].

\bibitem{AliAhmad:2025oli}
S.~Ali~Ahmad and M.S.~Klinger, \emph{{Extensions from within}},
  \href{https://arxiv.org/abs/2503.02944}{{\ttfamily 2503.02944}}.

\bibitem{Connes1973}
A.~Connes, \emph{Une classification des facteurs de type $\{\rm iii\}$},
  {\emph{Annales scientifiques de l'École Normale Supérieure} {\bfseries 6}
  (1973) 133}.

\bibitem{ConnesTakesakiflow}
A.~Connes and M.~Takesaki, \emph{{The flow of weights on factors of type ${\rm
  III}$}}, \href{https://doi.org/10.2748/tmj/1178240493}{\emph{Tohoku
  Mathematical Journal} {\bfseries 29} (1977) 473 }.

\bibitem{haagerup1979operatorII}
U.~Haagerup, \emph{{Operator valued weights in von Neumann algebras, II}},
  {\emph{Journal of Functional Analysis} {\bfseries 33} (1979) 339}.

\bibitem{vonNeumann1932}
J.~von Neumann, \emph{Proof of the quasi-ergodic hypothesis},
  \href{https://doi.org/10.1073/pnas.18.1.70}{\emph{Proceedings of the National
  Academy of Sciences} {\bfseries 18} (1932) 70}.

\bibitem{Lashkari:2019ixo}
N.~Lashkari, \emph{{Modular zero modes and sewing the states of QFT}},
  \href{https://doi.org/10.1007/JHEP04(2021)189}{\emph{JHEP} {\bfseries 21}
  (2020) 189} [\href{https://arxiv.org/abs/1911.11153}{{\ttfamily
  1911.11153}}].

\bibitem{Neumann1930}
J.v.~Neumann, \emph{Zur algebra der funktionaloperationen und theorie der
  normalen operatoren}, {\emph{Mathematische Annalen} {\bfseries 102} (1930)
  370}.

\bibitem{Gieres:2021ekc}
F.~Gieres, \emph{{Covariant canonical formulations of classical field
  theories}},
  \href{https://doi.org/10.21468/SciPostPhysLectNotes.77}{\emph{SciPost Phys.
  Lect. Notes} {\bfseries 77} (2023) 1}
  [\href{https://arxiv.org/abs/2109.07330}{{\ttfamily 2109.07330}}].

\bibitem{Hamilton:2006az}
A.~Hamilton, D.N.~Kabat, G.~Lifschytz and D.A.~Lowe, \emph{{Holographic
  representation of local bulk operators}},
  \href{https://doi.org/10.1103/PhysRevD.74.066009}{\emph{Phys. Rev. D}
  {\bfseries 74} (2006) 066009}
  [\href{https://arxiv.org/abs/hep-th/0606141}{{\ttfamily hep-th/0606141}}].

\bibitem{Hamilton:2006fh}
A.~Hamilton, D.N.~Kabat, G.~Lifschytz and D.A.~Lowe, \emph{{Local bulk
  operators in AdS/CFT: A Holographic description of the black hole interior}},
  \href{https://doi.org/10.1103/PhysRevD.75.106001}{\emph{Phys. Rev. D}
  {\bfseries 75} (2007) 106001}
  [\href{https://arxiv.org/abs/hep-th/0612053}{{\ttfamily hep-th/0612053}}].

\bibitem{Morrison:2014jha}
I.A.~Morrison, \emph{{Boundary-to-bulk maps for AdS causal wedges and the
  Reeh-Schlieder property in holography}},
  \href{https://doi.org/10.1007/JHEP05(2014)053}{\emph{JHEP} {\bfseries 05}
  (2014) 053} [\href{https://arxiv.org/abs/1403.3426}{{\ttfamily 1403.3426}}].

\bibitem{Bousso:2012mh}
R.~Bousso, B.~Freivogel, S.~Leichenauer, V.~Rosenhaus and C.~Zukowski,
  \emph{{Null Geodesics, Local CFT Operators and AdS/CFT for Subregions}},
  \href{https://doi.org/10.1103/PhysRevD.88.064057}{\emph{Phys. Rev. D}
  {\bfseries 88} (2013) 064057}
  [\href{https://arxiv.org/abs/1209.4641}{{\ttfamily 1209.4641}}].

\bibitem{Papadodimas:2012aq}
K.~Papadodimas and S.~Raju, \emph{{An Infalling Observer in AdS/CFT}},
  \href{https://doi.org/10.1007/JHEP10(2013)212}{\emph{JHEP} {\bfseries 10}
  (2013) 212} [\href{https://arxiv.org/abs/1211.6767}{{\ttfamily 1211.6767}}].

\bibitem{Faulkner:2017vdd}
T.~Faulkner and A.~Lewkowycz, \emph{{Bulk Locality from Modular Flow}},
  \href{https://doi.org/10.1007/JHEP07(2017)151}{\emph{JHEP} {\bfseries 07}
  (2017) 151} [\href{https://arxiv.org/abs/1704.05464}{{\ttfamily
  1704.05464}}].

\bibitem{Borchers:2000pv}
H.J.~Borchers, \emph{{On revolutionizing quantum field theory with Tomita's
  modular theory}}, \href{https://doi.org/10.1063/1.533323}{\emph{J. Math.
  Phys.} {\bfseries 41} (2000) 3604}.

\bibitem{Shenker:2013pqa}
S.H.~Shenker and D.~Stanford, \emph{{Black holes and the butterfly effect}},
  \href{https://doi.org/10.1007/JHEP03(2014)067}{\emph{JHEP} {\bfseries 03}
  (2014) 067} [\href{https://arxiv.org/abs/1306.0622}{{\ttfamily 1306.0622}}].

\bibitem{Maldacena:2015waa}
J.~Maldacena, S.H.~Shenker and D.~Stanford, \emph{{A bound on chaos}},
  \href{https://doi.org/10.1007/JHEP08(2016)106}{\emph{JHEP} {\bfseries 08}
  (2016) 106} [\href{https://arxiv.org/abs/1503.01409}{{\ttfamily
  1503.01409}}].

\bibitem{Cirafici:2024jdw}
M.~Cirafici, \emph{{On the nonequilibrium dynamics of gravitational algebras}},
  \href{https://doi.org/10.1088/1361-6382/ad85bf}{\emph{Class. Quant. Grav.}
  {\bfseries 41} (2024) 235006}
  [\href{https://arxiv.org/abs/2402.03939}{{\ttfamily 2402.03939}}].

\bibitem{Fewster:2024pur}
J.C.~Fewster, D.W.~Janssen, L.D.~Loveridge, K.~Rejzner and J.~Waldron,
  \emph{{Quantum Reference Frames, Measurement Schemes and the Type of Local
  Algebras in Quantum Field Theory}},
  \href{https://doi.org/10.1007/s00220-024-05180-7}{\emph{Commun. Math. Phys.}
  {\bfseries 406} (2025) 19}
  [\href{https://arxiv.org/abs/2403.11973}{{\ttfamily 2403.11973}}].

\bibitem{Klinger:2023tgi}
M.S.~Klinger and R.G.~Leigh, \emph{{Crossed products, extended phase spaces and
  the resolution of entanglement singularities}},
  \href{https://doi.org/10.1016/j.nuclphysb.2024.116453}{\emph{Nucl. Phys. B}
  {\bfseries 999} (2024) 116453}
  [\href{https://arxiv.org/abs/2306.09314}{{\ttfamily 2306.09314}}].

\bibitem{Klinger:2023auu}
M.S.~Klinger and R.G.~Leigh, \emph{{Crossed products, conditional expectations
  and constraint quantization}},
  \href{https://doi.org/10.1016/j.nuclphysb.2024.116622}{\emph{Nucl. Phys. B}
  {\bfseries 1006} (2024) 116622}
  [\href{https://arxiv.org/abs/2312.16678}{{\ttfamily 2312.16678}}].

\bibitem{Liu:2021kay}
S.~Liu and B.~Yoshida, \emph{{Soft thermodynamics of gravitational shock
  wave}}, \href{https://doi.org/10.1103/PhysRevD.105.026003}{\emph{Phys. Rev.
  D} {\bfseries 105} (2022) 026003}
  [\href{https://arxiv.org/abs/2104.13377}{{\ttfamily 2104.13377}}].

\bibitem{Hollands:2023abp}
S.~Hollands and A.~Ranallo, \emph{{Channel Divergences and Complexity in
  Algebraic QFT}},
  \href{https://doi.org/10.1007/s00220-023-04855-x}{\emph{Commun. Math. Phys.}
  {\bfseries 404} (2023) 927}
  [\href{https://arxiv.org/abs/2302.10013}{{\ttfamily 2302.10013}}].

\bibitem{Abramowitz}
M.~Abramowitz and I.A.~Stegun, \emph{Handbook of Mathematical Functions with
  Formulas, Graphs, and Mathematical Tables}, Dover, New York (1964).

\end{thebibliography}\endgroup
